\documentclass[reprint, amsmath, amssymb, aps, onecolumn, draft, notitlepage]{revtex4-1}

\usepackage{dcolumn}
\usepackage{bm}
\usepackage{bbm}
\usepackage{hyperref}
\usepackage{braket}
\usepackage{physics}
\usepackage[utf8]{inputenc}
\usepackage[english]{babel}
\usepackage{amsthm}
\usepackage{xcolor}

\newtheoremstyle{definition2}
{12pt}
{12pt}
{}
{17pt }
{\bfseries}
{:}
{.5em}
{}

\theoremstyle{definition2}
\newtheorem{prop}{Proposition}

\begin{document}

\title{Quantum geometric maps and their properties}

\author{Marco Finocchiaro}
\email{marco.finocchiaro@aei.mpg.de}
\affiliation{Max Planck Institute for Gravitational Physics (Albert Einstein Institute), Am Muehlenberg 1, D-14476 Potsdam-Golm, Germany, EU}
\affiliation{Institute for Physics, Humboldt-Universit\"at zu Berlin, Newtonstraße 15, 12489 Berlin, Germany, EU}

\author{Yoobin Jeong}
	\email{ybj0530@gmail.com}
	\affiliation{Arnold Sommerfeld Center for Theoretical Physics, \\ Ludwig-Maximilians-Universit\"at München \\ Theresienstrasse 37, 80333 M\"unchen, Germany}

\author{Daniele Oriti}
	\email{daniele.oriti@physik.lmu.de}
	\affiliation{Arnold Sommerfeld Center for Theoretical Physics, \\ Ludwig-Maximilians-Universit\"at München \\ Theresienstrasse 37, 80333 M\"unchen, Germany}
	
\date{\today}

\begin{abstract}
Quantum geometric maps, which relate $SU(2)$ spin networks and Lorentz covariant projected spin networks, are an important ingredient of spin foam models (and tensorial group field theories) for 4-dimensional quantum gravity. We give a general definition of such maps, that encompasses all current spin foam models, and we investigate their properties at such general level. We then specialize the definition to see how the precise implementation of simplicity constraints affects features of the quantum geometric maps in specific models.
\end{abstract}

\maketitle

\section{Introduction}
Spin foam models \cite{Baez1998,Baez:1999sr, Perez2003} are combinatorial and algebraic formulations of the dynamics of quantum spacetime, in a covariant path integral-like language, defined by the assignment of a quantum amplitude to 2-complexes labeled by representations of a Lie group. While the spin foam formalism has wider applicability, their interest for quantum gravity stems from two main features, common to all models studied in this context. The first is that spin foam models for quantum gravity are dual to lattice gravity path integrals \cite{Finocchiaro2019}, with the gravitational degrees of freedom discretized on the lattice (usually a simplicial complex) dual to the spin foam 2-complex (by Poincar\'{e} duality). The second is that boundary data of spin foam amplitudes, for appropriate models, define spin networks, i.e. the same fundamental structures of the quantum geometry of canonical Loop Quantum Gravity (LQG) \cite{rovelli2004, Thiemann:2007zz, rovellividotto2014}. Spin foam models are also the Feynman amplitudes of (tensorial) group field theories \cite{Oriti:2011jm, krajewski2012group, oriti2014group}, a generalization to higher dimensions of matrix models for 2-dimensional quantum gravity, which also provide completion of the spin foam formalism embedding it into a sum over complexes (and discrete topologies).

The basic strategy for defining interesting spin foam models for 4-dimensional quantum gravity has been to work in a simplicial setting. A simplicial complex plays the role of discrete counterpart of the spacetime manifold and its Poincar\'{e} dual 2-skeleton provides the combinatorial structure over which the spin foam model is defined. The assignment of algebraic data (and the choice of their corresponding quantum amplitude) to such complex is, in turn, dictated by a description of the simplicial geometry (starting from that of their building blocks, i.e. tetrahedra for spatial slices and boundaries, and 4-simplices for the 4-dimensional simplicial complex) in terms of a phase space given by the cotangent bundle of a group manifold. There are two different, but classically equivalent, characterizations of a single tetrahedron in the simplicial geometry \cite{Baez1999, Pereira:2010wzm}. The first is to assign a vector $b_f^I\in\mathbb{R}^{3,1}$ (whose norm is proportional to area of the face) to each face $f$ of the tetrahedron $\tau$ and impose constraint $x_I \cdot b^I=0$ for a timelike vector $x_\tau$ such that the tetrahedron lies in a spacelike hypersurface. To ensure that four vectors form a tetrahedron, the closure constraint $\sum_{f\in\tau}b_f^I$ is imposed. This characterization, due to the constraints, allows us to associate a vector $b_f$ with the Lie algebra $\mathfrak{su(2)}$. Thus space $\mathfrak{su(2)}^{\times 4}$ provides a space of a single tetrahedron which can also be seen as a cotangent bundle $(T^*SU(2))^{\times 4}$. The conjugate variables in $SU(2)^{\times 4}$ represents the parallel transport of a discrete connection from the center of the tetrahedron to the center of its boundary faces. In this picture, the dual graph of this tetrahedron becomes a spin network with a 4-valent vertex. One can construct, by properly defining the inner product and imposing the constraint, the Hilbert space $L^2(SU(2)^{\times4})$ for the single quantum tetrahedron. The second characterization is to employ a bivector $B^{IJ}_f \in \wedge^2\mathbb{R}^{3,1}\cong \mathfrak{sl(2,c)}$ which is close $\sum_{f\in\tau}B_f^{IJ}$ and constrained by  simplicity constraints $x_I\cdot(*B)^I=0$ for a tetrahedron. This second characterization suggests the Hilbert space for the single quantum tetrahedron is $L^2(SL(2,\mathbb{C})^{\times 4}, \mathcal{H}^+)$ where $\mathcal{H}^+$ is a set of timelike vector $x_\tau$ (with $x^0_\tau>0$) under the simplicity constraint at the quantum level. Two quantum descriptions of a tetrahedron and full simplicial geometry can be related via a map that we call here \lq quantum geometric map'.

The second construction as a continuum counterpart in the observation that the topological BF action yields the Palatini first order gravity action under simplicity constraints \cite{doi:10.1063/1.523215}. Thus one can define a path integral of the BF action on a discrete manifold and impose the simplicity constraint at the quantum level \cite{Pietri_1999, Freidel1998SpinFM}. The resulting regularized (by the means of the discretization) path integral of the action provides the partition function of the quantum gravity model and the amplitude for a simplicial complex, which can then be recast, equivalently, in the spin foam language \cite{Finocchiaro2019}. A different choice of quantum simplicity constraint (for the same classical simplicity constraint), configuration space, and gauge group produces different spin foam models having different amplitudes, while most of spin foam models for 4-dimensional gravity carry the Lorentz group $SL(2,\mathbb{C})$ (or Spin(4) for the Riemannian theory) as a gauge group.

Such spin foam models are often referred to as a covariant counterpart of Loop Quantum Gravity, in the sense that they are expected to provide a way to describe the dynamics of spin network states which has proven difficult to control by a quantum Hamiltonian constraint in the canonical framework.
However, describing the LQG dynamics using the spin foam formalism requires relating two different quantum states of two theories as the LQG Hilbert space is formed by the $SU(2)$ spin networks and the boundary Hilbert space for the spin foam model is spanned naturally by the projected spin networks \cite{Livine2002,  Alexandrov:2002br}, which in fact arise straightforwardly from the covariant, constrained BF description of classical and quantum geometry. 

The type of quantum variables characterizing them, in particular, the irreducible representation assigned on each edge, and of course the underlying simplicial geometric description, suggest that the $SU(2)$ spin network can be embedded into the Lorentz-based projected spin networks, and vice versa that the covariant projected spin network can be projected down to the $SU(2)$ spin network \cite{PhysRevD.82.064044}. In fact, one could expect to be able to translate back and forth between these two descriptions of the quantum geometry of these models. The \lq quantum geometric maps' mentioned before take part of this translation between two descriptions.

These quantum geometric maps, therefore, allow to formulate spin foam models and their boundary states in two alternative manners. One, in terms of explicitly covariant data taken from the $SL(2,\mathbb{C})$ or $Spin(4)$ group, and the other in terms of their rotation 3-dimensional rotation subgroup only, with the 4-dimensional covariance properties of the models (and the states) encoded in the dynamical amplitudes. The same choice is obviously available also in the group field theory context \cite{oriti2011microscopic, 2006gr.qc.....7032O}, which provides a complete definition of the same spin foam models by embedding them in a field theory context, generalizing matrix and tensor models \cite{osti_10110632, Gurau_2012, gurau2019notes}, and allowing both a precise definition of the sum over spin foam complexes and more direct access to non-perturbative dynamics and collective physics of spin network degrees of freedom \cite{oriti2015group, Carrozza_2016, baloitcha2020flowing}. 
The models defined following the same quantization strategies of classical structures (in particular, the simplicity constraints) but using these two types of data are not equivalent, in general. The corresponding Hilbert space of states is different, and so is, in general, the quantum dynamics. 
The precise relation between models differing only by this choice of boundary states, and a precise characterization of their similarities and differences, however, depends on the detailed properties of the map relating the two formulations. That is, it depends on the \lq quantum geometric map' one has employed.  The general structure and properties of such maps are the object of our analysis.
{}

A specific type of quantum geometric map has been constructed by M. Dupuis and E. Livine, through the convolution of character functions, and including an undetermined factor (which is not irrelevant for the properties of the map) in \cite{PhysRevD.82.064044}, and adapted to the EPRL-FK imposition of the simplicity constraints. The analysis of the properties of this map has raised questions concerning the incompatibility between the two reasonable requirements of isometry and of the embedding map from $SU(2)$ spin networks to covariant ones being the inverse (suitably intended) of a projection map, that would be the most intuitive way of understanding how $SU(2)$ spin networks arise from covariant ones. In particular, one would like to understand if this incompatibility is a generic feature of quantum geometric maps, whether it depends on other properties having been assumed for the same maps, or whether it follows from the choice of quantum imposition of the simplicity constraints (and thus may not arise in other spin foam models, based on different imposition strategies and characterized by different encoding of the same constraints as restrictions on the group-theoretic data). 

In this article, we give first of all a general definition of the quantum geometric maps, that encompasses all spin foam models in this \lq constrained BF' class (i.e. all imposition strategies for the simplicity constraints). Next, we investigate some properties of the embedding and projection maps at such general level, as well as the compatibility between different properties. Finally, we specialize the definition to see how the precise implementation of simplicity constraints affects features of the quantum geometric maps, and what happens in specific spin foam models.

\section{Hilbert spaces and gauge symmetries}

As we already discussed, embedding and projection maps are mappings between the $SU(2)$ cylindrical functions and projected cylindrical functions, associated to graphs (which can be taken to be embedded in a topological manifold (in canonical LQG) or defining a simplicial complex (in usual spin foam models and in group field theory). These are two different, but presumably geometrically equivalent (once also the dynamics is implemented) representations of a 3-geometry in the spin foam, loop quantum gravity and tensorial group field theory formalisms. 

The full Hilbert space in which they are included differ in the mentioned approaches, in the way states associated to different graphs are related \cite{Oriti2014}, but as long as restricts consideration only to the Hilbert space associated to a given graph, the Hilbert spaces in these formalisms coincide. Since we are only concerned, here, with the relation between $SU(2)$ and covariant states, we restrict our attention to what happens for given fixed graph. 

In this section, we construct the Hilbert spaces of different types of cylindrical functions defined on the directed graph $\Gamma$ with $E$ edges and $V$ vertices.

First consider the $SU(2)$ cylindrical functions. The Hilbert space $\mathcal{K}^{SU(2)}$ of the cylindrical functions $\phi$ on $E$ copies of $SU(2)$ is
\begin{equation}
\mathcal{K}^{SU(2)} = \{ \phi : SU(2)^E \longrightarrow \mathbb{C} \text{ $\lvert$ }\lVert \phi \lVert^2 < \infty\}
\end{equation}
where the $L^2$ norm is induced by the inner product
\begin{equation}
\langle \phi, \tilde{\phi} \rangle_{SU(2)}=\int_{SU(2)} [dg_e]^E \overline{\phi(g_e)}\tilde{\phi}(g_e)
\end{equation}
with the Haar measure $dg_e$ on the group manifold $SU(2)$. Here $\phi(g_e)\equiv \phi(g_1, ..., g_E)$.

The group action on the state on the graph $\Gamma$ is defined as the multiplication of the group elements associated to the edges by a distinct group element associated to every vertices of the graph, the multiplications being from the left or from the right depending on whether the vertex is a source or a target for the given edge. The $SU(2)$ cylindrical function is said to be $SU(2)$ invariant if it is invariant under this gauge transformation:
\begin{equation}
\phi(g_e) = \phi(h_{s(e)}g_{e}h_{t(e)}^{-1}), \text{ } \forall h_v \in SU(2)
\end{equation}
where $s(e)$ is the source vertex for  the edge $e$ incidents and $t(e)$ is the vertex where the same edge $e$ terminates. The $SU(2)$ gauge invariant functions form the \lq kinematical Hilbert space\rq, that we label $\mathcal{K}^{SU(2)}_{kin}$.  A special case of the above is when the underlying graph has (out-going) open edges, i.e. edges which end on uni-valent vertices, where we assume no action of the gauge group; for example, assuming a graph withh all $E$ edges being open, the gauge invariance reads
\begin{equation}
\phi(g_e) = \phi(h_{s(1)}g_{1}, ..., h_{s(E)}g_{E}), \text{ } \forall h_v \in SU(2)
\end{equation}
In the opposite situation with all edges being in-coming, one would have instead
\begin{equation}
\phi(g_e) = \phi(g_{1}h_{t(1)}^{-1}, ..., g_{E}h_{t(E)}^{-1}), \text{ } \forall h_v \in SU(2)
\end{equation}
%The gauge invariance of the functions defined on the graph with several open edges can be easily generalized. 
This gauge invariant property corresponds to the Gauss constraint of LQG.
Upon Peter-Weyl decomposition, these cylindrical functions expand in the standard spin network basis, each spin network state being characterized by an assignment of an irreducible representation of $SU(2)$ on each edge with angular momentum projection index at each end of the edge, contracted by invariant tensors (intertwiners) associated to the vertices of the graph.

Now let us consider the covariant counterpart of the $SU(2)$ cylindrical functions, on both Riemannian and Lorentzian signatures, i.e. the ones that serve as a starting point for defining quantum gravity states in four dimensions, via the constrained BF strategy. A basis for the relevant covariant states is formed by so-called projected spin networks \cite{Livine2002}. A projected spin network is a directed graph with a (time-like, in the Lorentzian case) unit vector $x_v$ on each vertex, an irreducible representation of $G$ ($Spin(4)$ for the Riemannian theory and $SL(2, \mathbb{C})$ for the Lorentzian theory) and two irreducible representations of the stabilizer group $SU_{x_v}(2)$ on each edge (one for each vertex connected by the edge) and an $SU_{x_v}(2)$ intertwiner on each vertex, resulting from the same invariance we described for standard spin networks, now with respect to the $x_v$-dependent $SU(2)$ subgroup of the group $G$. The corresponding cylindrical function is a function on $E$ copies of $G$ and $V$ copies of the homogeneous space $Q$ where
\begin{equation}
Q=
\begin{cases}
Spin(4)/SU(2) \cong S^3 \text{ for the Riemannian case}\\
SL(2,\mathbb{C})/SU(2) \cong \mathcal{H}^+ \text{ for the Lorentzian case.}
\end{cases}
\end{equation}
Here $S^3=\{x\in \mathbb{R}^4 \text{ $\lvert$ } \lVert x\rVert^2=1 \}$ is a 3-sphere and
$\mathcal{H}^+=\{x\in\mathbb{R}^{3,1}\text{ $\lvert$ } x^0 >0 \text{ and } \lVert x\rVert^2=1\}$ is a hyperboloid \cite{Livine2002, PhysRevD.82.064044}.
We now discuss the two resulting Hilbert spaces in some more detail, separating the Riemannian and the Lorentzian cases.

\subsection{Riemannian theory}

The space of the cylindrical functions we consider is the Hilbert space $\mathcal{K}^{Spin(4)}$
\begin{equation}
\mathcal{K}^{Spin(4)} = \{ \psi : Spin(4)^E\times (S^3)^V \longrightarrow \mathbb{C} \text{ $\lvert$ }\lVert \psi \lVert^2_{Spin(4)} < \infty\}
\end{equation}
where the norm $\lVert \cdot \lVert_{Spin(4)}$ is induced by the inner product
\begin{equation}\label{innerproduct_Euclidean}
\langle \psi, \tilde{\psi} \rangle_{Spin(4)}=\int_{S^3}  [dx_v]^V \int_{Spin(4)} [dG_e]^E  \overline{\psi(G_e,x_v)}\tilde{\psi}(G_e,x_v)
\end{equation}
with the Haar measure $dG_e$ on the group manifold $Spin(4)$ and the Lebesgue measure $dx_v$ on the homogeneous space $Spin(4)/SU(2)\cong S^3$.

Just like the $SU(2)$ case, we can require an additional symmetry condition to be satisfied by the cylindrical functions of interest. The relevant cylindrical functions are $Spin(4)$ invariant if
\begin{equation}
\psi(G_e, x_v) = \psi(H_{s(e)}G_{e}H_{t(e)}^{-1}, H\rhd x_v), \forall H_v \in Spin(4).
\end{equation}
While the group action on the group elements labelling the edges is defined similarly in the Riemannian and Lorentzian cases, the group action on the normal vector $x_v$ should be adapted to the case at hand. In the Riemannian case, the action is 
\begin{equation}
H \rhd x_v = h^+ x_v (h^-)^{-1}
\end{equation}
where $H=(h^+, h^-)$, $h^\pm \in SU(2)$ using the isomorphism $Spin(4)\cong SU(2) \times SU(2)$ and the $x_v \in S^3$ is seen as an $SU(2)$ group element under the identification \cite{Freidel2008}
\begin{equation}
x_v=(x^0, x^1, x^2, x^3) \sim
\begin{pmatrix}
x^0+ix^3 & x^2 + ix^1 \\
-x^2+i x^1 & x^0-ix^3 
\end{pmatrix}.
\end{equation}
The $Spin(4)$ action so defined is a 4-dimensional rotation of the unit 4-vector $x_v$. Some rotations do not change the vector and form the stabilizer group, which is the $SU(2)_{x_v}$ group previously mentioned. The $Spin(4)$ invariance so defined induces another symmetry under the action of the stabilizer group of $x_v$:
\begin{equation}
\psi(G_e, x_v) = \psi(h_{s(e)}G_{e}h_{t(e)}^{-1}, x_v), \forall h_v \in SU_{x_v}(2) \subset Spin(4).
\end{equation}
A $Spin(4)$ invariant function can also be obtained, of course, by acting with a projector $P_{inv}$ on any cylindrical function $\psi\in\mathcal{K}^{Spin(4)}$
\begin{equation}
\big(P_{inv}\psi\big)(g_e, x_v)=\int_{Spin(4)} [dH_e]^E\psi(H_{s(e)}g_eH_{t(e)}^{-1}, H \rhd x_v).
\end{equation}
The cylindrical functions with $Spin(4)$ invariance form a kinematical Hilbert space $\mathcal{K}^{Spin(4)}_{kin}$ which is equipped with the same inner product as (\ref{innerproduct_Euclidean}). Note that this inner product on the kinematical Hilbert space carries a redundant integral due to the $Spin(4)$ invariance
\begin{align}\label{inner_Euclidean2}
\langle \psi, \tilde{\psi} \rangle_{Spin(4)}&=\int_{S^3}  [dx_v]^V \int_{Spin(4)} [dG_e]^E  \overline{\psi(G_e, x_v)}\tilde{\psi}(G_e, x_v) \\
&=\int_{S^3}  [dx_v]^V \int_{Spin(4)} [dG_e]^E  \overline{\psi(H_{s(e)}(k)G_eH_{t(e)}^{-1}(k), \mathbbm{1})}\tilde{\psi}(H_{s(e)}(k)G_eH_{t(e)}^{-1}(k), \mathbbm{1})\nonumber\\
&=\int_{S^3}  [dx_v]^V \int_{Spin(4)} [dG_e]^E  \overline{\psi(G_e, \mathbbm{1})}\tilde{\psi}(G_e, \mathbbm{1})\nonumber
\end{align}
where $H_v(x_v) \rhd \mathbbm{1} = x_v$ and the invariant properties of the Haar measure are used. The expression (\ref{inner_Euclidean2}) provides the inner product for the gauge-fixed function $\varphi(G_e):=\psi(G_e, x_v=\mathbbm{1})$ for the $Spin(4)$ invariant functions. Note that the gauge-fixed functions lose the full $Spin(4)$ invariance and have only induced $SU_{x_v}(2)$ invariance. The $x_v$ integrals result in a factor equal to the volume of the homogeneous space, which we take to be normalized to one.
The \lq projected\rq nature of the canonical basis of cylindrical functions, which would be otherwise simply a generalization of standard spin networks from $SU(2)$ to $Spin(4)$, arise when the $Spin(4)$ representation associated to each edge is further decomposed into a canonical basis of representation functions for the $SU(2)_{x_v}$ subgroup at each end point of the edge, incident to the vertex to which the vector $x_v$ is associated \cite{Livine2002}.  

So far, these covariant cylindrical functions have no gravitational or geometric characterization. In fact, they can be seen as quantum states for a topological BF theory discretized on the graph $\Gamma$. Spin foam models for 4-dimensional gravity are constructed, as discussed, from such discrete BF theory and imposing the simplicity constraints, which in both continuum and discrete classical formulations turn the topological theory into the geometric (first order) gravity theory, at the quantum level. The quantum implementation of such constraints is a subtle matter (and a main focus of attention of the spin foam community over the last years) and different ways of imposing the simplicity constraints yield different models. 

In general terms, the imposition of the simplicity constraints is a mapping $S^{\omega}$
\begin{eqnarray}\label{mapS}
S^{\omega}: \mathcal{K}^{Spin(4)} &\longrightarrow& \mathcal{K}^{Spin(4)}\\
\psi &\longmapsto& \psi^\omega \nonumber.
\end{eqnarray}
which in practice, affects the expansion of the resulting cylindrical functions in terms of irreducible representations of the group $Spin(4)$ and, once this is performed, in terms of the representations of the stabilizer groups $SU(2)_{x_v}$. We characterize this modified expansion in representations in terms of a coefficient $\omega$ (closely related to the so-called fusion coefficients), which thus encodes the precise implementation of the simplicity constraints. To see more explicitly the results of the imposition of the simplicity constraints in the representation space, one first needs to decompose the cylindrical functions into the representation functions. For the cylindrical functions $\psi$ in the kinematical Hilbert spaces, this looks as follows:
\begin{equation}
\psi(G_e, x_v)=\sum_{J_i,M_i,N_i} [\psi(x_v)]^{J_e}_{M_eN_e}\prod_{i=e}^E D^{J_e}_{M_eN_e}(G_e)
\end{equation}
by the Peter-Weyl decomposition \cite{MartinDussaud2019APO}. $D^J$ is the representation function of the irreducible representation $J$ of $Spin(4)$. The $\psi^J$ are the modes of the cylindrical function. Then, upon the imposition of the simplicity constraints the same expansion is modified as
\begin{equation}
\psi^{\omega}(G_e, x_v)=\sum_{J_e,M_e,N_e, j_e} [\psi(x_v)]^{J_e}_{M_eN_e}\prod_{e=1}^E \Big(D^{J_e}_{M_eN_e}(G_e)\omega(J_e, j_e, \gamma)\Big)
\end{equation}
where the coefficient $\omega$ constrains the further expansion of the representation $J$ into irreducible representations of the subgroup $SU(2)_{x_v}$.
Here $j$ is the irreducible representations of $SU(2)$ and the parameter $\gamma$ is the Immirzi parameter \cite{Holst:1995pc,ENGLE2008136,Perez:2005pm}. As, said, the coefficient $\omega$ restricts the decomposition:
\begin{equation}
\mathcal{H}^{J} = \bigoplus_{|j^+-j^-|}^{j^++j^-}\mathcal{H}^j 
\end{equation}
where $J=(j^+, j^-)$. We refer to \cite{Finocchiaro2019} for more details.

\subsection{Lorentzian theory}
In the Lorentzian case, the definition of the relevant Hilbert spaces is entirely analogous, but of course extra care should be taken due to the non-compactness of the Lorentz group.

The relevant space of the cylindrical functions is the Hilbert space $\mathcal{K}^{SL(2,\mathbb{C})}$
\begin{equation}
\mathcal{K}^{SL(2,\mathbb{C})} = \{ \psi : SL(2,\mathbb{C})^E\times (\mathcal{H}^+)^V \longrightarrow \mathbb{C} \text{ $\lvert$ }\lVert \psi \lVert^2_{SL(2,\mathbb{C})} < \infty\}
\end{equation}
where the norm $\lVert \cdot \lVert_{SL(2,\mathbb{C})}$ is induced by the inner product
\begin{equation}\label{innerproduct_Lorentzian}
\langle \psi, \tilde{\psi} \rangle_{SL(2,\mathbb{C})}=\int_{\mathcal{H}^+} [dx_v]^V \int_{SL(2,\mathbb{C})} [dG_e]^E \overline{\psi(G_e,x_v)}\tilde{\psi}(G_e,x_v).
\end{equation}
with the Haar measure $dG_e$ on the group manifold $SL(2, \mathbb{C})$ and the Lebesgue measure $dx_v$ on the homogeneous space $SL(2, \mathbb{C})/SU(2)\cong \mathcal{H}^+$.

The cylindrical function is $SL(2,\mathbb{C})$ invariant if
\begin{equation}
\psi(G_e, x_v) = \psi(H_{s(e)}G_{e}H_{t(e)}^{-1}, H\rhd x_v), \forall H_v \in SL(2,\mathbb{C}).
\end{equation}
where the group action on the vertex vectors is
\begin{equation}
H \rhd x_v = H x_v H^{^\dagger}
\end{equation}
where the vector $x_v$ can again be seen as an $SU(2)$ group element under the identification \cite{Freidel2008}
\begin{equation}
x_v=(x^0, x^1, x^2, x^3) \sim
\begin{pmatrix}
x^0+x^3 & x^2 - ix^1 \\
x^2+i x^1 & x^0-x^3 
\end{pmatrix}.
\end{equation}
Just like in the Riemannian case, this $SL(2, \mathbb{C})$ action is a (3,1)-dimensional Lorentz rotation of the timelike 4-vector whose stabilizer group is again an $SU_{x_v}(2)$ group. The $SL(2, \mathbb{C})$ invariance thus induces another symmetry under the action of the stabilizer group of $x_v$:
\begin{equation}
\psi(G_e, x_v) = \psi(h_{s(e)}G_{e}h_{t(e)}^{-1}, x_v), \forall h_v \in SU_{x_v}(2) \subset SL(2, \mathbb{C}).
\end{equation}
Also in this Lorentzian case, an $SL(2,\mathbb{C})$ invariant function can be obtained by acting with a \lq projector\rq $P_{inv}$ on a generic cylindrical function $\psi \in \mathcal{K}^{SL(2,\mathbb{C})}$
\begin{equation}
\big(P_{inv}\psi\big)(g_e, x_v) = \int_{SL(2, \mathbb{C})} [dH_e]^E \psi(H_{s(e)}g_eH^{-1}_{t(e)}, H\rhd x_v) \qquad .
\end{equation}
However, this is a formal definition only, since this \lq projector operator\rq would produce immediately a divergence when acting more than once, due to the non-compact domain of integration. A (rather straightforward) regularization procedure will therefore be needed whenever this construction is used.
The Lorentz invariant cylindrical functions form a (kinematical) Hilbert space $\mathcal{K}^{SL(2, \mathbb{C})}_{kin}$ which equips the same inner product as (\ref{innerproduct_Lorentzian}). We note again that the inner product carries a divergent integral due to the $SL(2, \mathbb{C})$ invariance
\begin{align}
\langle \psi, \tilde{\psi} \rangle_{SL(2, \mathbb{C})}&=\int_{\mathcal{H}^+}  [dx_v]^V \int_{SL(2, \mathbb{C})} [dG_e]^E  \overline{\psi(G_e, x_v)}\tilde{\psi}(G_e, x_v)\\
&=\int_{\mathcal{H}^+}  [dx_v]^V \int_{SL(2, \mathbb{C})} [dG_e]^E  \overline{\psi(H_{s(e)}(x_v)G_eH_{t(e)}^{-1}(x_v), \mathbbm{1})}\tilde{\psi}(H_{s(e)}(x_v)G_eH_{t(e)}^{-1}(x_v), \mathbbm{1})\nonumber\\
&=\int_{\mathcal{H}^+}  [dx_v]^V \int_{SL(2, \mathbb{C})} [dG_e]^E  \overline{\psi(G_e, \mathbbm{1})}\tilde{\psi}(G_e, \mathbbm{1})\nonumber
\end{align}
where $H_{v}(x_v)\rhd\mathbbm{1}=x_v$, due to the non-compactness of $\mathcal{H}^+$. However, this divergence is not physical and the inner product can be regularized by simply dropping the integral that corresponds to the volume of the homogeneous space $\mathcal{H}^+$. The resulting expression provides the inner product for the gauge-fixed function $\varphi(G_e):=\psi(G_e, x_v=\mathbbm{1})$ for the $SL(2, \mathbb{C})$ invariant functions. Obviously, this gauge-fixed function loses the full $SL(2, \mathbb{C})$ invariance, only the induced $SU_{x_v}(2)$ invariance remains.

Imposing the simplicity constraint in the Lorentzian theory is not much different from in the Riemannian theory. The constraint is imposed by a mapping $S^{\omega}$
\begin{eqnarray}
S^{\omega}: \mathcal{K}^{SL(2, \mathbb{C})} &\longrightarrow& \mathcal{K}^{SL(2, \mathbb{C})}\\
\psi &\longmapsto& \psi^\omega \nonumber.
\end{eqnarray}
Upon the Plancherel decomposition \cite{PhysRevD.82.064044}, the cylindrical function is represented as
\begin{align}
\psi(G_e, x_v)=\sum_{\substack{a_e, j_e, l_e,\\ m_e, n_e}}\int \prod_{e=1}^E &\Big(\mu(\rho_e, a_e) d\rho_e\Big) \psi^{(\rho_e, a_e)}_{j_em_el_en_e}(x_v) \prod_{i=1}^E D^{(\rho_e, a_e)}_{j_em_el_en_e}(G_e)
\end{align}
where $a\in \mathbb{N}/2$ and $\rho \in \mathbb{R}$. Here $D^{(\rho, a)}$ is the representation function for the irreducible (unitary) representation $(\rho, a)$ of $SL(2, \mathbb{C})$ of the principal series and $\mu(\rho,a)=(\rho^2+a^2)$ is the Plancherel measure. The coefficient $\psi^{(\rho,a)}$ are the modes of the cylindrical function. Then the simplicity constraint imposition results in a modification of this expansion as
\begin{equation}
\psi^{\omega}(G_e, x_v)=\sum_{\substack{a_e, j_e, l_e, k_e, \\ m_e, n_e}}\int \prod_{e=1}^E \Big(\mu(\rho_e, a_e) d\rho_e\Big) \psi^{(\rho_e, a_e)}_{j_em_el_en_e}(x_v)\prod_{e=1}^E \Big(D^{(\rho_e, a_e)}_{j_em_el_en_e}(G_e)\omega((\rho_e, a_e), k_e, \gamma_e)\Big).
\end{equation}
The $j$, $k$, and $l$ label elements in the canonical basis for the irreducible representations of $SU(2)$. This way of expressing the restriction uses the fact that the representation space of $SL(2, \mathbb{C})$ can be expressed as the direct sum of the representation spaces of $SU(2)$:
\begin{equation}
\mathcal{H}^{(\rho, a)} = \bigoplus_{j\in a+\mathbb{N}}\mathcal{H}^j.
\end{equation}
%Therefore the implementation of the simplicity constraint in at the spin representation implies choosing a specific subspace for each representation of $SL(2, \mathbb{C})$.

We refer to the spin foam literature (e.g. \cite{Perez:2012wv}) for more details.

\section{embedding and projection maps}\label{section:embedding and projection maps}

Being equipped with the different Hilbert spaces we are interested in, we can now define the embedding and projection maps relating the two, and thus relating the $SU(2)$ cylindrical functions and the covariant (projected) cylindrical functions. The maps can be defined at different levels depending on the symmetries being imposed. In particular, we will focus our attention on the properties of the maps, especially, injectivity, isometry, and, indeed, implementation of the $SU(2)$ and $SL(2,\mathbb{C})$ (or $Spin(4)$) symmetries. 

Let us spend a few words to clarify the importance of these properties. The injectivity guarantees that different $SU(2)$ spin network states correspond to different covariant boundary states on the spin foam. If the embedding map is not injective, then a non-trivial $SU(2)$ spin network state could be mapped to the zero state, producing thus a highly degenerate spin foam amplitude. In other words, it would simply be impossible to capture the spin foam dynamics in terms of its effect on $SU(2)$ spin networks, and thus no canonical LQG interpretation could possibly be given to it. The isometry property amounts to the requirement that the map between $SU(2)$ and covariant quantum states preserves the inner products, including of course preserving their norm. 
This is a stronger requirement than injectivity and would ensure that matrix elements of kinematical observables are preserved in the two formulations, and that, so to speak, no information about the quantum spin foam dynamics gets lost when passing from one to the other. Finally, the states in the kinematical Hilbert spaces have symmetries so the maps between the state spaces could be required to preserve those symmetries, to ensure their correct implementation in the spin foam dynamics regardless of the formulation being chosen for its boundary data.

The embedding and projection maps can be defined between the Hilbert spaces without gauge symmetries or at the gauge invariant level. 

First, let us consider the maps between Hilbert spaces without gauge symmetries. 

The embedding map $K'$ is a map from $\mathcal{K}^{SU(2)}$ to $\mathcal{K}^{G}$:
\begin{eqnarray}
K' : \mathcal{K}^{SU(2)} &\longrightarrow \mathcal{K}^{G}\\
\phi &\longmapsto \psi' \nonumber
\end{eqnarray}
The most general form of the embedding map can be written with the help of the integral kernel $K'(G_i, g_i, x_v)$:
\begin{equation}
\psi(G_e, x_v) = \int_{SU(2)} [dg_e]^E K'(G_e, g_e, x_v) \phi(g_e)
\end{equation}
where $\phi(g_e) = \phi (g_1, ..., g_E)$ and the same applies to $K'(G_e, g_e, x_v)$ and $\psi(G_e, x_v)$.

The projection map $P'$ is a map from  $\mathcal{K}^{G}$ to $\mathcal{K}^{SU(2)}$:
\begin{eqnarray}
P' : \mathcal{K}^{G} &\longrightarrow \mathcal{K}^{SU(2)}\\
\psi' &\longmapsto \phi \nonumber
\end{eqnarray}
Just as the embedding map, the most general form of the projection map can be written in terms of the integral kernel $P'(G_e, g_e, x_v)$:
\begin{equation}
\phi(g_e) = \int_Q [dx_v]^V \int_{G}  [dG_e]^{E} P'(G_e, g_e, x_v) \psi(G_e, x_v).
\end{equation}

One can also define the embedding and projection maps between Hilbert spaces that incorporate the mentioned gauge symmetries. The corresponding embedding map is
\begin{eqnarray}
K: \mathcal{K}^{SU(2)}_{kin} &\longrightarrow \mathcal{K}^{G}_{kin}\\
\phi &\longmapsto \psi \nonumber
\end{eqnarray}
and the integral kernel is defined by
\begin{equation}
\psi(G_e, x_v) = \int_{SU(2)} [dg_e]^{E} K(G_e, g_e, x_v) \phi(g_e).
\end{equation}

Moreover, a different embedding map incorporating also the simplicity constraints can be defined as
\begin{eqnarray}
K^\omega: \mathcal{K}^{SU(2)}_{kin} &\longrightarrow \mathcal{K}^{G}_{kin}\\
\phi &\longmapsto \psi^{\omega} \nonumber
\end{eqnarray}
This embedding map is nothing but $K^\omega = S \circ K$. The integral kernel is defined by
\begin{equation}
\psi^\omega(G_e, x_v) = \int_{SU(2)} [dg_e]^E K^\omega(G_e, g_e, x_v) \phi(g_e).
\end{equation}

Similarly, the projection map is defined as
\begin{eqnarray}
P: \mathcal{K}^{G}_{kin}& \longrightarrow \mathcal{K}^{SU(2)}_{kin}\\
\psi& \longmapsto \phi. \nonumber
\end{eqnarray}
Its integral kernel is defined by
\begin{equation}
\phi(g_e)=\int_{Q}[dx_v]^V\int_{G}[dG_e]^E P(G_e, g_e, x_v)\psi(G_e, x_v).
\end{equation}

Having the definitions, let us now consider the properties of the embedding map and the projection map.

\subsection{Riemannian theory}

\subsubsection{Injectivity}

The injectivity of the embedding map $K'$ would amount to
\begin{equation}\label{injectivity_K}
\text{if }K'(\phi)=0 \text{ then } \phi=0.
\end{equation}
The correspondent statement applies to the embedding maps $K$ and $K^\omega$. In practice, in order to analyse it and to verify whether it is satisfied by specific constructions, it is useful to write the condition in terms of the expression of the integral kernels in terms of group representations. 

The integral kernel of $K'$ is decomposed as
\begin{equation}
K'(G_i, g_i, x_v) \equiv \sum_{\substack{J_i ,M_i, N_i,  \\j_i, m_i, n_i}} [K'(x_v)]^{J_i j_i}_{M_iN_i m_in_i} \prod_{i=1}^E\Big(D^{J_i}_{M_iN_i}(G_i)\overline{D^{j_i}_{m_in_i}(g_i)}\Big)
\end{equation}
under the Peter-Weyl decomposition. This gives the \lq spin\rq representation of the integral kernel $K'$ as:
\begin{align}\label{eq:Rspinrep}
[K'(x_v)]^{J_ij_i}_{M_iN_im_in_i}=\prod_{i=1}^E&\Big(d_{J_i}d_{j_i}\Big)\int_{Spin(4)}\prod_{i=1}^E dg_i K(G_i, g_i, x_v) \prod_{i=1}^E\Big( \overline{D^{J_i}_{M_iN_i}(G)}D^{j_i}_{m_in_i}(g_i)\Big).
\end{align}
using the orthogonality relations of the matrix elements of the representations \cite{MartinDussaud2019APO}
\begin{eqnarray}
d_j\int_{SU(2)} dg D^{j}_{mn}(g)\overline{D^{l}_{pq}(g)}&= \delta^{jl}\delta_{mp}\delta_{nq},\label{su2orthorel}\\
d_J\int_{Spin(4)} dG D^{J}_{MN}(G)\overline{D^{L}_{PQ}(G)}&= \delta^{JL}\delta_{MP}\delta_{NQ}.
\end{eqnarray}
Having the spin representation of the integral kernel at hand, one can write the injectivity condition in terms of it: if
\begin{equation}
\sum_{j_i, m_i, n_i}\int_{S^3}\prod_{v=1}^V dx_v [K'(x_v)]^{J_ij_i}_{M_iN_im_in_i} \prod_{i=1}^E\frac{1}{d_{j_i}}\phi^{l_i}_{m_in_i}=0 \text{ $\forall$}J_i, M_i, N_i, \text{ implies } \phi^{j_i}_{p_iq_i}=0 \text{ $\forall$} l_i, p_i, q_i
\end{equation}
then $K'$ is injective. 

Similarly, the integral kernel of $K^\omega$ is decomposed as
\begin{equation}
K^\omega(G_i, g_i, x_v) \equiv \sum_{\substack{J_i ,M_i, N_i, \\ j_i, m_i, n_i}} [K^\omega(x_v)]^{J_i j_i}_{M_iN_i m_in_i} \prod_{i=1}^E\Big(D^{J_i}_{M_iN_i}(G_i)\overline{D^{j_i}_{m_in_i}(g_i)}\Big).
\end{equation}
And its spin representation is
\begin{equation}
[K^\omega(x_v)]^{J_ij_i}_{M_iN_im_in_i}=\prod_{i=1}^E\Big(d_{J_i}d_{j_i}\Big)\int_{Spin(4)} [dg_i]^E K(G_i, g_i, x_v) \prod_{i=1}^E\Big( \overline{D^{J_i}_{M_iN_i}(G)}D^{j_i}_{m_in_i}(g_i) \omega(J_i, j_i, \gamma) \Big).
\end{equation}
The injectivity condition of $K^\omega$, in terms of such spin representation, is then that: if
\begin{equation}\label{eq:injectivity}
\sum_{j_i, m_i, n_i} [K^\omega(x_v)]^{J_ij_i}_{M_iN_im_in_i}\prod_{i=1}^E\frac{1}{d_{j_i}}\phi^{j_i}_{m_in_i}=0 \text{ $\forall$} J_i, M_i, N_i, \text{ implies } \phi^{l_i}_{p_iq_i}=0 \text{ $\forall$} l_i, p_i, q_i
\end{equation}
then $K^\omega$ is injective.

\subsubsection{Isometry}

The isometry of the embedding map $K'$ is the condition that
\begin{equation}\label{isometry_K}
\langle \phi, \tilde{\phi} \rangle_{SU(2)} = \langle K'(\phi), K'(\tilde{\phi}) \rangle_{Spin(4)}.
\end{equation}

If $K'$ is isometric then its integral kernels satisfies
\begin{equation}\label{isometry_K_kernel}
\int_{S^3} [dx_v]^V \int_{Spin(4)}  [dG_e]^E K'(G_e, g_e, x_v)\overline{K'(G_e, h_e, x_v)}= \prod_{e=1}^E \delta_{SU(2)} (g_eh_e^{-1}).
\end{equation}
In fact, the isometry of any map defined in a normed vector space implies the injectivity of the same map (the proof is recalled in section \ref{section:Relations}). The converse is not true in general, therefore any non-injective map is non-isometric.

The isometry condition (\ref{isometry_K}) can be written in terms of the spin representation as
\begin{equation}\label{Isometry_K_spin}
\sum_{I_i, M_i, N_i} \int_{S^3} [dx_v]^V \prod_{i=1}^E\frac{1}{d_{I_i}}[K'(x_v)]^{I_ij_i}_{M_iN_im_in_i}\overline{[K'(x_v)]^{I_ik_i}_{M_iN_ip_iq_i}} = \prod_{i=1}^E\Big(d^{j_i}\delta_{m_ip_i}\delta_{n_iq_i} \delta^{j_ik_i}\Big).
\end{equation}

Moving then to the more interesting quantum geometric context, the embedding map $K^\omega$ is isometric if
\begin{equation}\label{isometry_K2}
\langle \phi, \tilde{\phi} \rangle_{SU(2)} = \langle K^\omega(\phi), K^\omega(\tilde{\phi}) \rangle_{G}.
\end{equation}
If the map is isometric then its integral kernel satisfies
\begin{align}\label{isometry_kernel_Riemannian}
\int_Q [dx_v]^V \int_{Spin(4)} [dG_e]^E K^\omega(G_e, g_e, x_v)\overline{K^\omega(G_e, h_e, x_v)}= \prod_{e=1}^E \delta_{SU(2)} (g_eh_e^{-1}).
\end{align}
The spin representation of (\ref{isometry_kernel_Riemannian}) is
\begin{equation}\label{Isometry_K_spin2}
\sum_{I_i, M_i, N_i}\int_Q [dx_v]^V \prod_{i=1}^E\frac{1}{d_{I_i}}[K^\omega(x_v)]^{I_ij_i}_{M_iN_im_in_i}\overline{[K^\omega(x_v)]^{I_ik_i}_{M_iN_ip_iq_i}}
=\prod_{i=1}^E\Big(d^{j_i}\delta_{m_ip_i}\delta_{n_iq_i} \delta^{j_ik_i}\Big).
\end{equation}

It is immediate to verify that the isometry condition for the embedding map $K$, corresponding to the non-geometric, topological case, is obtained as a special case when all $\omega$ are set to be the identity.

\subsubsection{Miscellaneous}
We now state a number of other properties of the embedding and projection maps that turn out to be useful in the following analysis, and are also of more general interest.

When the restriction of the image of embedding map $K$ from $Spin(4)$ to $SU_{\mathbbm{1}}(2)$ yields the mapped function, $\psi((g_i, g_i), \mathbbm{1})=K(\phi)(g_i)=\phi(g_i)$, the integral kernel of the embedding map satisfies
\begin{equation}
K((g_e, g_e), h_e, \mathbbm{1}) = \prod_{e=1}^E\delta_{SU(2)}(g_eh_e^{-1}).
\end{equation}
The spin representation of the condition above is
\begin{equation}
\sum_{m_i^{\pm}, n_i^{\pm}}[K(\mathbbm{1})]^{j^+_ij^-_il_i}_{m^+_in^+_im^-_in^-_im_in_i}\prod_{i=1}^E\Big(C^{j^+_ij^-_ij_i}_{m^+_im^-_ip_i}\overline{C^{j^+_ij^-_ij_i}_{n^+_in^-_iq_i}}\Big)=\prod_{i=1}^E\Big(d_{j_i}\delta^{j_il_i}\delta_{p_im_i}\delta_{q_in_i}\Big).
\end{equation}

If the embedding map $K'$ maps $SU(2)$ invariant function to $SU_{x_v}(2)$ invariant function defined on a closed graph then the integral kernel satisfies
\begin{equation}\label{symtranslation}
K'(h_{s(e)} G_e h_{t(e)}^{-1}, h_e, x_v) = K'(G_e, h_{s(e)}^{-1} h_eh_{t(e)}, x_v).
\end{equation}
Of course the $SU_{x_v}(2)$ invariance does not ensure the $Spin(4)$ invariance. 

This symmetry \lq translation\rq is equivalent to the invariance of the integral kernel under simultaneous left diagonal $SU_{x_{t(e)}}(2) (\subset Spin(4))$ action and right diagonal $SU_{x_{s(e)}}(2) (\subset Spin(4))$ action on $Spin(4)$ and $SU(2)$ action on $SU(2)$ itself. The left invariance can be implemented by a projector
\begin{equation}
P_{inv}^{L}(K')(G_e, g_e, x_v) = \int_{SU(2)} dh K'( (hg^+_e, x_{s(e)}^{-1}hx_{s(e)}g^-_e), hg, x_v).
\end{equation}
The right invariance can be imposed similarly. 

These invariances imply, in turn, that the group representation of the integral kernel is a convolution of $Spin(4)$ and $SU(2)$ characters. In the spin representation, each invariance intertwines $Spin(4)$ and $SU(2)$ representations providing in such a way that the expression can be given in terms of rotated Clebsch-Gordon coefficients
\begin{equation}
[K(x_v)]^{j^+_ij^-_ij_i}_{m^+_in^+_im^-_in^-_im_in_i}\propto \prod_{i=1}^E\Big(\overline{C^{j^+_ij^-_ij_i}_{m^+_im^-_im_i}(x_{s(e)})}C^{j^+_ij^-_ij_i}_{n^+_in^-_in_i}(x_{t(e)})\Big),
\end{equation}
where the rotated CG is defined by
\begin{equation}
C^{j^+j^-j}_{m^+m^-m}(x_v)\overline{C^{j^+j^-j}_{n^+n^-n}(x_v)}=d_j\int_{SU(2)} dg D^{j^+}_{m^+n^+}(g)D^{j^-}_{m^-n^-}(x_v^{-1}gx_v)\overline{D^{j}_{mn}(g)}
\end{equation}
This shows that the symmetry requirement fully determines the integral kernel of the map up to some proportionality weights.

On a closed graph, the embedding map $K^\omega$  maps an $SU(2)$ invariant cylindrical function to a $Spin(4)$ invariant function in $\mathcal{K}^{Spin(4)}_{kin}$, thus the integral kernel satisfies
\begin{equation}
K(G_e, g_e, x_v)=K(H_{s(e)}G_eH_{t(e)}^{-1}, g_e, H_{v}\rhd x_v)
\end{equation}
for any $H_{v}\in Spin(4)$. Moreover, the $SU(2)$ symmetry is translated into the induced $SU_{x_v}(2)$ symmetry and the integral kernel satisfies (\ref{symtranslation}).

The projection map $P$ maps instead a $Spin(4)$ invariant function to an $SU(2)$ invariant functions. Therefore the integral kernel satisfies
\begin{equation}\label{projspin4}
P(G_e, g_e, x_v)=P(H_{s(e)}G_eH_{t(e)}^{-1}, g_e, H_{v}\rhd x_v)
\end{equation}
and
\begin{equation}\label{projsu2}
P(G_e, g_e, x_v)=P(G_e, h_{s(e)}g_eh_{t(e)}^{-1}, x_v)
\end{equation}
for any $H_{v}\in Spin(4)$ and $h_{x_v}\in SU_{x_v}(2)$.

\subsection{Lorentzian theory}

\subsubsection{Injectivity}
The injectivity condition for the embedding map in the Lorentzian theory is the same as (\ref{injectivity_K}). In order to write the injective condition in terms of group representations, let us first introduce the \lq spin\rq representation of the integral kernels. The integral kernel of $K'$ is decomposed as
\begin{equation}
K'(G_i, g_i, x_v) \equiv \sum_{\substack{a_i, j_i ,l_i, \\ m_i,  n_i, p_i, q_i}}\int \prod_{i=1}^E\big(\mu(\rho_i, a_i)d\rho_i\big) [K'(x_v)]^{(\rho_i, a_i) k_i}_{j_im_i l_in_ip_i q_i} \prod_{i=1}^E\Big(D^{(\rho_i, a_i)}_{j_im_il_in_i}(G_i)\overline{D^{j_i}_{p_iq_i}(g_i)}\Big)
\end{equation}
via the Plancherel decomposition for the $SL(2,\mathbb{C})$ part and the Peter-Weyl decomposition for the $SU(2)$ part. The equation can be inverted to obtain the spin representation of $K'$:
\begin{equation}
[K'(x_v)]^{(\rho_i, a_i)k_i}_{j_im_il_in_ip_iq_i}=\prod_{i=1}^E\frac{d_{j_i}}{\mu(\rho_i, a_i)} \int_{SL(2, \mathbb{C})} \prod_{i=1}^E dG_i  
 K'(G_i, g_i, x_v) \prod_{i=1}^E\Big(\overline{D^{(\rho_i, a_i)}_{j_im_il_in_i}(G_i)}D^{k_i}_{p_iq_i}(g_i) \Big).
\end{equation}
using the orthogonality relations of the matrix elements of the representations \cite{MartinDussaud2019APO}
\begin{equation}
\int_{SL(2, \mathbb{C})} dG D^{(\rho, a)}_{jmln}(G)\overline{D^{(\rho', a')}_{j'm'l'n'}(G)}= \frac{1}{\mu(\rho, a)}\delta(\rho-\rho')\delta_{jj'}\delta_{ll'}\delta_{mm'}\delta_{nn'}\delta_{aa'}
\end{equation}
and (\ref{su2orthorel}). 

The injectivity condition in terms of the spin representation reads
\begin{equation}
\sum_{k_i,p_i,q_i}\int_{\mathcal{H}^+}\prod_{v=1}^Vdx_v[K'(x_v)]^{(\rho_i,a_i)k_i}_{j_im_il_in_ip_iq_i}\prod_{i=1}^E\frac{1}{d_{k_i}}\phi^{k_i}_{p_iq_i}=0 \text{ $\forall$} \rho_i, a_i, j_i, m_i, l_i, n_i \text{ implies }\phi^{k'_i}_{p'_iq'_i}=0 \text{ $\forall$} k'_i, p'_i, q'_i .
\end{equation}
Moreover, the integral kernel of $K^\omega$ is decomposed as
\begin{equation}
K^\omega(G_i,  g_i, x_v)=\sum_{\substack{a_i, k_i, l_i,\\ m_i, n_i, p_i, q_i}}\int \prod_{i=1}^4 \Big(\mu(\rho_i, a_i)d\rho_i\Big)[K^\omega(x_v)]^{(\rho_i, a_i)k_i}_{j_im_il_in_ip_iq_i} \prod_{i=1}^4\Big(D^{(\rho_i, a_i)}_{j_im_il_in_i}(G_i)\overline{D^{k_i}_{p_iq_i}(g_i)}\Big),
\end{equation}
with
\begin{equation}
[K^\omega(x_v)]^{(\rho_i, a_i)k_i}_{j_im_il_in_ip_iq_i}=\prod_{i=1}^Ed_{k_i} \int_{SL(2, \mathbb{C})} [dG_i]^E K(G_i, g_i, x_v)
\prod_{i=1}^E\Big(\overline{D^{(\rho_i, a_i)}_{j_im_il_in_i}(G_i)}D^{k_i}_{p_iq_i}(g_i)\omega((\rho_i, a_i), k_i, \gamma) \Big).
\end{equation}
The injectivity condition in the spin representation is
\begin{equation}\label{injectivity_spin_Lorentz}
\sum_{k_i,p_i,q_i}[K(x_v)]^{(\rho_i,a_i)k_i}_{j_im_il_in_ip_iq_i}\prod_{i=1}^E\Big(\frac{1}{d_{k_i}}\tilde{\omega}((\rho_i,a_i), k_i ,\gamma)\Big)\phi^{k_i}_{p_iq_i}=0 \text{ $\forall$} \rho_i, a_i,  j_i, l_i, m_i, n_i \text{ implies }\phi^{k'_i}_{p'_iq'_i}=0 \text{ $\forall$} k_i', p'_i, q'_i.
\end{equation}
Here $\tilde{\omega}$ is $\omega$ which is replaced the delta distribution to the Kronecker delta.

\subsubsection{Isometry}
The isometry condition for the embedding map $K'$ is expressed as
\begin{equation}\label{isometry_K3}
\langle \phi, \tilde{\phi} \rangle_{SU(2)} = \langle K'(\phi), K'(\tilde{\phi}) \rangle_{SL(2, \mathbb{C})}.
\end{equation}
Again, the isometry of the map implies its injectivity. 

One can also write the isometry condition (\ref{isometry_K3}) in terms of the integral kernel of an injective embedding map as
\begin{equation}
\int_{\mathcal{H}^+} [dx_v]^V \int_{SL(2, \mathbb{C})}  [dG_e]^E K'(G_e, g_e, x_v)\overline{K'(G_e, h_e, x_v)}
= \prod_{e=1}^E \delta_{SU(2)} (g_eh_e^{-1}).
\end{equation}

The spin representation of the isometry condition reads
\begin{equation}
\sum_{\substack{a_i, j_i, l_i, \\ m_i, n_i}}\int_{\mathcal{H}^+} [dx_v]^V \int \prod_{i=1}^E \Big( \mu(\rho_{i},a_{i}) d\rho_i \Big) [K'(x_v)]^{(\rho_i, a_i)k_i}_{j_im_il_in_ip_iq_i} \overline{[K'(x_v)]^{(\rho_i, a_i)k_i'}_{j_im_il_in_ip_i'q_i'}} = \prod_{i=1}^E\Big(d^{k_i}\delta_{p_ip_i'}\delta_{q_iq_i'} \delta^{k_ik_i'}\Big) \qquad .
\end{equation}

The isometry condition for the embedding map $K^\omega$ is the same (\ref{isometry_K}). The equation can be expressed in terms of the integral kernels
\begin{equation}
\int_{\mathcal{H}^+}[dx_v]^V\int_{SL(2,\mathbb{C})}[dG_e]^E K^\omega(G_e, h_e, x_v)\overline{K^\omega(G_e, g_e, x_v)} = \prod_{e=1}^E \delta(g_eh_e^{-1}).
\end{equation}
The isometry condition in terms of the integral kernel expressed in group representations reads
\begin{equation}\label{isometry_K_spin_Lorentzian}
\sum_{\substack{a_i, j_i, l_i, \\ m_i, n_i}} \int_{\mathcal{H^+}}[dx_v]^V \int \prod_{i=1}^E\Big(\mu(\rho_{i},a_{i})d\rho_i\Big)[K^\omega (x_v)]^{(\rho_i, a_i)k_i}_{j_im_il_in_ip_iq_i} \overline{[K^\omega(x_v)]^{(\rho_i, a_i)k_i'}_{j_im_il_in_ip_i'q_i'}} 
= \prod_{i=1}^E\Big(d^{k_i}\delta_{p_ip_i'}\delta_{q_iq_i'} \delta^{k_ik_i'}\Big).
\end{equation}
Like in the Riemannian case, the isometry condition on the embedding map $K$ can be obtained from this latter one, as a special case, when all $\omega$ are set to the identity.

\subsubsection{Miscellaneous}
Also for the Lorentzian theory, we close this section with some additional useful properties of the embedding and projection maps.

When the restriction of the image of embedding map from $SL(2, \mathbb{C})$ to $SU(2)$ yields the original function, we have
\begin{equation}
\int_{\mathcal{H}^+}[dx_v]^V K(g_e, h_e, x_v) = \prod_{e=1}^E\delta_{SU(2)}(g_eh_e^{-1}).
\end{equation}

For the embedding map $K'$ on a closed graph, if the $SU(2)$ invariance is translated into the induced $SU(2)_x$ invariance then
\begin{equation}\label{symmetrytranslation}
K'(h_{s(e)} G_e h_{t(e)}^{-1}, g_e, x_v) = K'(G_e, h_{s(e)}^{-1} g_e h_{t(e)}, x_v).
\end{equation}
Again, obviously the $SU(2)$ invariance does not ensure $SL(2, \mathbb{C})$ invariance.  

This symmetry translation is equivalent to the invariance of the integral kernel under simultaneous left diagonal $SU_{x_{t(e)}}(2) (\subset Spin(4))$ action and right diagonal $SU_{x_{s(e)}}(2) (\subset Spin(4))$ action on $Spin(4)$, and the $SU(2)$ action on $SU(2)$ itself. 

These invariances imply that the group representation of the integral kernel is a convolution of $SL(2,\mathbb{C})$ and $SU(2)$ characters. 

In the spin representation, each invariance intertwines $SL(2, \mathbb{C})$ and $SU(2)$ representations implying the expression
\begin{equation}
[K(x_v)]^{(\rho_i, a_i)k_i}_{j_im_il_in_ip_iq_i}\propto 
\begin{cases}
\prod\limits_{i = 1}^{E} \Big(\delta_{m_ip_i}\delta_{n_iq_i}\delta^{j_ik_i}\delta^{j_il_i}\Big) &\text{ if }\in a+\mathbb{N}\\
0 &\text{ if }\notin a+\mathbb{N}
\end{cases}
\end{equation}
Again, as in the Riemannian case, the symmetry requirement determines the embedding map up to weight factor.

The embedding map $K$ maps an $SU(2)$ function to a $SL(2, \mathbb{C})$ invariant function in $\mathcal{K}^{SL(2, \mathbb{C})}_{kin}$, thus the integral kernel satisfies
\begin{equation}
K(G_e, g_e, x_v)=K(H_{s(e)}G_eH_{t(e)}^{-1}, g_e, H_{x_v}\rhd x_v)
\end{equation}
for any $H_{x_v}\in SL(2, \mathbb{C})$. The $SU(2)$ symmetry is translated into the induced $SU_{x_v}(2)$ symmetry and the integral kernel satisfies (\ref{symmetrytranslation}).

The projection map in the Lorentzian theory satisfies the relations (\ref{projspin4}) and (\ref{projsu2}) as well.

\section{Cylindrical functions for a 4-valent (open) vertex}

%The gauge symmetries of functions in $\mathcal{K}^{SU(2)}_{kin}$ and $\mathcal{K}^{G}_{kin}$ give certain structure to the functions. 

Let us restrict our interest to a direct graph which has only a single vertex which has four out-going edges with open ends for the simplicity. This case is the one needed for immediate application to spin foam models and to the group field theory formalism. This restriction is also sufficient to investigate the properties of the embedding and projection maps in the general graph case, since the full graph can be obtained by gluing several single vertices with open edges, and the properties of the embedding and projection map of the full graph are determined by the properties of each vertex building block.

\subsection{$SU(2)$ states}
Any function $\phi \in \mathcal{K}^{SU(2)}_{kin}$ defined on such graph satisfies
\begin{equation}
\phi(g_i)= \phi(hg_i), \forall h \in SU(2)
\end{equation}
which leads to 
\begin{equation}
\phi^{j_i}_{m_in_i}=  \sum_{b} A^{j_ib}_{n_i}  \overline{(\mathcal{I})}^{j_ib}_{m_i}
\end{equation}
where $A$ is an arbitrary tensor,  $(\mathcal{I})^{j_ib}_{m_i} \equiv (\mathcal{I})^{j_1j_2j_3j_4b}_{m_1m_2m_3m_4}$ is an 4-valent intertwiner \cite{Finocchiaro2019}, and the spin representation is defined by
\begin{equation}
\phi(g_i)= \sum_{j_i, m_i, n_i}\phi^{j_i}_{m_in_i} \prod_{i=1}^4 D^{j_i}_{m_in_i}(g_i) \qquad.
\end{equation}

\subsection{Riemannian theory}
Any projected cylindrical function $\psi \in \mathcal{K}^{Spin(4)}_{kin}$ defined on the graph has the induced $SU_{x_v}(2)$ invariance
\begin{equation}\label{induced su2}
\psi(G_i, x_v)= \psi(hG_i, x_v), \forall h \in SU(2)
\end{equation}
which yields \cite{Finocchiaro2019},
\begin{equation}
[\psi(x_v)]^{j^+_ij^-_i}_{m^+_im^-_in^+_in^-_i}=\sum_{j_i, m_i^{(\pm)}, n_i^{(\pm)}, l } \big[B(x_v)\big]^{j^+_ij^-_i l}_{n^+_in^-_i}\overline{C^{j^+_ij^-_ij_i}_{m^+_im^-_im_i}(x_v)}\overline{(\mathcal{I})}^{j_il}_{m_i, x_v}
\end{equation}
where $B$ is an arbitrary tensor and $(\mathcal{I})^{j_ib}_{m_i, x_v}$ is an 4-valent intertwiner of $SU_{x_v}(2)$. The spin representation of the cylindrical function is defined by the following decomposition
\begin{equation}
\psi ((g^+_i, g^-_i), x_v) = \sum_{j^{\pm}_i, m_i^\pm, n_i^\pm } [\psi(x_v)]^{j^+_i j^-_i}_{m^+_i n^+_i m^-_i n^-_i} \prod_{i=1}^4 \Big(D^{j^+_i}_{m^+_in^+_i}(g^+_i)D^{j^-_i}_{m^-_in^-_i}(g^-_i)\Big) \qquad .
\end{equation}
One can write the embedding map in terms of the spin representation as
\begin{equation}
\sum_{l_i,a_i, l}\big[B(x_v)\big]^{j^+_ij^-_il}_{n^+_in^-_i}\overline{C^{j^+_ij^-_il_i}_{m^+_im^-_ia_i}(x_v)}\overline{(\mathcal{I})}^{l_il}_{a_i, x_v}=\sum_{j_i, k, m_i, n_i}\prod_{i=1}^4 \frac{1}{d_{j_i}} [K(x_v)]^{j^+_ij^-_ij_i}_{m^+_im^-_in^+_in^-_im_in_i}\overline{(\mathcal{I})}^{j_ik}_{m_i}A^{j_ik}_{n_i} \qquad .
\end{equation}
A simple calculation then gives
\begin{equation}
[B(x_v)]^{j^+_ij^-_il}_{n^+_in^-_i} =\sum_{k, j_i, n_i} [F(x_v)]^{lkj^+_ij^-_ij_i}_{n_i^+n^-_in_i} A^{j_ik}_{n_i}
\end{equation}
where
\begin{align}
[F(x_v)]^{lkj^+_ij^-_ij_i}_{n_i^+n^-_in_i}=\sum_{l_i, a_i,  m^{(\pm)}_i}\prod_{i=1}^4 \frac{1}{d_{j_i}} [K(x_v)]^{j^+_ij^-_ij_i}_{m^+_im^-_in^+_in^-_im_in_i} C^{j^+_ij^-_il_i}_{m^+_im^-_ia_i}(x_v) (\mathcal{I})^{l_il}_{a_i, x_v}\overline{(\mathcal{I})}^{j_ik}_{m_i} \qquad.
\end{align}
The orthogonality of 4-valent intertwiners \cite{Finocchiaro2019}
\begin{equation}
\sum_{m_i}(\mathcal{I})^{j_il}_{m_i, (x_v)}\overline{(\mathcal{I})}_{m_i, (x_v)}^{j_ik}=\delta^{lk}
\end{equation}
and the orthogonality of the CG coefficients can be used then to invert the 4-valent intertwiners and the CG coefficients. Since $A$ and $B$ are arbitrary tensors, so is $F$. The spin representation of $K$ can be expressed as
\begin{equation}
[K(x_v)]^{j^+_ij^-_ij_i}_{m^+_im^-_in^+_in^-_im_in_i}  = \sum_{k, l_i, l, a_i}\prod_{i=1}^4 \Big(d_{j_i}\overline{C^{j^+_ij^-_il_i}_{m^+_im^-_ia_i}(x_v)}\Big)(\mathcal{I})^{j_ik}_{m_i}\overline{(\mathcal{I})}^{l_il}_{a_i, x_v} [F(x_v)]^{lkj^+_ij^-_ij_i}_{n_i^+n^-_in_i},
\end{equation}
for a tensor $F$. 

When the simplicity constraints are imposed, the spin representation of $K^\omega$ reads
\begin{equation}\label{K}
[K^\omega(x_v)]^{j^+_ij^-_ij_i}_{m^+_im^-_in^+_in^-_im_in_i}= \sum_{k, l_i, l, a_i} \prod_{i=1}^4 \Big(d_{j_i}\overline{C^{j^+_ij^-_il_i}_{m^+_im^-_ia_i}(x_v)}\Big)(\mathcal{I})^{j_ik}_{m_i}\overline{(\mathcal{I})}^{l_il}_{a_i, x_v} [F(x_v)]^{lkj^+_ij^-_ij_i}_{n_i^+n^-_in_i}(\omega) \quad ,
\end{equation}
where
\begin{equation}
[F(x_v)]^{lkj^+_ij^-_ij_i}_{n_i^+n^-_in_i}(\omega)\equiv [F(x_v)]^{lkj^+_ij^-_ij_i}_{n_i^+n^-_in_i} \omega(j^+_i, j^-_i, j_i, \gamma) \qquad .
\end{equation}

The injectivity and the isometry of the embedding map will be the focus of the later discussion.

The injectivity condition has been given in (\ref{eq:injectivity}) but it can also be written in terms of the $F$ factor: 
\begin{equation}\label{injF}
\sum_{j_i, n_i, b}[F(x_v)]^{lbj_i^+j_i^-j_i}_{n_i^+n^-_in_i}\omega(j^+_i, j^-_i, j_i, \gamma)A^{j_ib}_{n_i}=0, \text{ $\forall$} j^+_i, j^-_i, l, n^+_i, n^-_i  \text{ implies }A^{j_ic}_{n_i}=0 \text{ $\forall$ } j_i, n_i, c \qquad .
\end{equation}
The isometry condition (\ref{Isometry_K_spin2}) in terms of $F$ is
\begin{equation}\label{isoF}
\sum_{\substack{x_i, j_i^{\pm}, \\ x,  n_i^{\pm}}}\prod_{i=1}^4\Big(\frac{d_{j_i}}{d_{j^+_i}d_{j^-_i}}\Big)[F(x_v)]^{xy j^+_i j^-_ix_i}_{n^+_i n^-_i n_i}(\omega)\overline{[F(x_v)]^{xz j^+_i j^-_ix_i}_{n^+_i n^-_i q_i}(\omega)}=\prod_{i=1}^4\delta_{n_iq_i} \delta^{yz} \qquad .
\end{equation}

\subsection{Lorentzian theory}
The induced $SU_{x_v}(2)$ invariance (\ref{induced su2}) yields
\begin{equation}
[\psi(x_v)]^{(\rho_i, a_i)}_{j_im_il_in_i}=\sum_k [B(x_v)]^{(p_i, a_i)j_ik}_{l_in_i}\overline{(\mathcal{I})}^{j_ik}_{m_i, x_v} \qquad .
\end{equation}
Here we use the fact that $SL(2,\mathbb{C})$ matrix element reduces to a $SU(2)$ matrix element when it is evaluated on elements of an $SU(2)$ subgroup of $SL(2,\mathbb{C})$:
\begin{equation}
D^{(\rho,a)}_{jmln}(H)=\delta_{jl}D^j_{mn}(H) \text{ for } H \in SU(2)\subset SL(2,\mathbb{C}) \qquad .
\end{equation}
The spin representation of the same expression is defined by the decomposition \cite{PhysRevD.82.064044}
\begin{equation}
\psi (G_i, x_v) = \sum_{\substack{a_i, j_i, l_i,\\ m_i, n_i}}\int \prod_{i=1}^4\Big(\mu(\rho_i,a_i) d\rho_i\Big) [\psi(x_v)]^{(\rho_i, a_i)}_{j_im_il_in_i} D^{(\rho_i, a_i)}_{j_im_il_in_i}(G_i) \qquad .
\end{equation}

The embedding map $K$ in the spin representation reads
\begin{equation}
\sum_{k}[B(x_v)]^{(\rho_i, a_i)j_ik}_{l_in_i}\overline{(\mathcal{I})}^{j_ik}_{m_i, x_v}=\sum_{\substack{k_i, l,\\ p_i, q_i}}\prod_{i=1}^4 \frac{1}{d_{k_i}} [K(x_v)]^{(\rho_i, a_i)k_i}_{j_im_il_in_ip_iq_i} A^{k_il}_{q_i}\overline{(\mathcal{I})}^{k_il}_{p_i} \qquad .
\end{equation}
By inverting the intertwiner function, one obtains
\begin{equation}\label{Bfactor}
[B(x_v)]^{(\rho_i, a_i)j_ik}_{l_in_i} =  \sum_{k_i, l, q_i}[F(x_v)]^{(\rho_i, a_i)j_ikk_il}_{l_in_iq_i} A^{k_il}_{q_i} \quad ,
\end{equation}
where $F$ is
\begin{equation}
[F(x_v)]^{(\rho_i, a_i)j_ikk_il}_{l_in_iq_i} = \sum_{m_i, p_i} \prod_{i=1}^4\frac{1}{d_{k_i}} [K(x_v)]^{(\rho_i, a_i)k_i}_{j_im_il_in_ip_iq_i} (\mathcal{I})^{j_ik}_{m_i, x_v}\overline{(\mathcal{I})}^{k_il}_{p_i} \qquad .
\end{equation}
Since $A$ and $B$ are arbitrary tensors, $F$ is also an arbitrary tensor. The spin representation of $K$ looks then as
\begin{equation}
K^{(\rho_i, a_i)k_i}_{j_im_il_in_ip_iq_i} = \sum_{k, l}\prod_{i=1}^4 d_{k_i}\overline{(\mathcal{I})}^{j_ik}_{m_i, x_v}(\mathcal{I})^{k_il}_{p_i} [F(x_v)]^{(\rho_i, a_i)j_ikk_il}_{l_in_iq_i},
\end{equation}
for a tensor $F$. When the simplicity constraints are imposed, the spin representation of $K^\omega$ reads
\begin{equation}\label{K3}
[K^\omega]^{(\rho_i, a_i)k_i}_{j_im_il_in_ip_iq_i} = \sum_{k,l}\prod_{i=1}^4 d_{k_i}\overline{(\mathcal{I})}^{j_ik}_{m_i, x_v}(\mathcal{I})^{k_il}_{p_i} [F(x_v)]^{(\rho_i, a_i)j_ikk_il}_{l_in_iq_i}(\omega) \quad ,
\end{equation}
where
\begin{equation}
[F(x_v)]^{(\rho_i, a_i)j_ikk_il}_{l_in_iq_i}(\omega)\equiv [F(x_v)]^{(\rho_i, a_i)j_ikk_il}_{l_in_iq_i} \omega((\rho_i, a_i), k_i, \gamma) \qquad .
\end{equation}
The embedding map $K$ is then entirely characterized only by the form of $F$, with the other factors entering its expression being fixed by the symmetry requirements.

\noindent The injectivity and isometry conditions on the maps can then be reduced to conditions on this function $F$ only.
The injectiviy condition (\ref{injectivity_spin_Lorentz}) in terms of the $F$ function is the condition that
\begin{equation}
\sum_{j_i, k_i,l,q_i}[F(x_v)]^{(\rho_i, a_i)j_ikk_il}_{l_in_iq_i}\tilde{\omega}((\rho_i, a_i), j_i, \gamma)A^{k_il}_{q_i}=0 \text{ $\forall$}\rho_i, a_i, n_i, k \text{ implies } A^{k_il}_{q_i}= 0 \text{ $\forall$} k_i, l, q_i \quad.
\end{equation}
The isometry condition (\ref{isometry_K_spin_Lorentzian}) in terms of the $F$ is instead the condition that
\begin{align}\label{iso3}
\sum_{a_i, j_i, l_i, k} \int\prod_{i=1}^4\Big(\mu(\rho_i, a_i)d\rho_i\Big)\prod_{i=1}^4\Big(d_{k_i}d_{k'_i}\Big)[F(x_v)]^{(\rho_i,a_i)j_ikk_il}_{l_in_iq_i}(\omega)\overline{[F(x_v)]^{(\rho_i,a_i)j_ikk_i'l'}_{l_in_iq_i'}(\omega)}=\sum_{j_i, p_i}\prod_{i=1}^4\Big(\frac{\delta_{q_iq_i'}}{d_{j_i}}\Big)\overline{(\mathcal{I})}^{k_il}_{p_i}(\mathcal{I})^{k_i'l'}_{p_i} \quad .
\end{align}

\section{Relations among properties of embedding and projection maps}\label{section:Relations}

We have investigated the conditions required for the maps to have the properties we are interested in, without assuming any specific form for the maps (thus without focusing on any specific spin foam model), except some basic symmetry requirement. Now we consider the compatibility between some desirable properties of the embedding and projection maps, including of course injectivity and isometry, and remaining at the same level of generality.

\

Let us start by considering the inverse of the embedding map.

If $K^\omega$ is injective, then there exists an inverse map \\
\begin{equation}
\tilde{P}: \text{Im}(K^\omega)\longrightarrow \mathcal{K}_{kin}^{SU(2)}
\end{equation} where $\tilde{P}\circ K^\omega = id_{SU(2)}$

A priori, this $\tilde{P}$ has nothing to do with the projection map. However, we can choose $\tilde{P}$ as the restriction of the projection map $P\vert_{\text{Im($K^\omega$)}}=\tilde{P}$.

Then, a few propositions can be easily proven.

\prop{$\tilde{P}$ is injective.}
\begin{proof}
For any $\psi_1^{\omega}\neq\psi_2^{\omega}$, there exists $\phi_1\neq\phi_2$ such that $K^\omega(\phi_1)\neq K^\omega(\phi_2)$. Since $\tilde{P}(\psi_i^{\omega}) = \tilde{P}(K^\omega(\phi_i))=\phi_i$, $\tilde{P}(\psi_1^{\omega})\neq \tilde{P}(\psi_2^{\omega})$. Therefore $\tilde{P}$ is injective.
\end{proof}

\prop{$K^\omega\circ \tilde{P}=id_{G}\rvert_{\text{Im}(K^\omega)}$.}
\begin{proof}
For any $\psi^{\omega}\in \text{Im}(K^\omega)$, there exists $\phi$ such that $\psi^{\omega}=K^{\omega}(\phi)$. Thus, $\tilde{P}(\psi^\omega)=\tilde{P}(K^\omega(\phi))$. By mapping it with $K^\omega$, $(K^\omega \circ \tilde{P})(\psi^\omega)=K^\omega ((\tilde{P}\circ K^\omega)(\phi))=K^\omega(\phi)=\psi^{\omega}$ which implies that $K^\omega \circ \tilde{P} = id_{G}\rvert_{\text{Im}(K^\omega)}$.
\end{proof}

\prop{If $K^\omega$ is isometric then it is injective.}
\begin{proof}
Suppose $K^\omega$ is isometric. For any $\phi \in \mathcal{K}^{SU(2)}_{kin}$ such that $K^\omega(\phi)=0$, $0=\langle K^\omega(\phi), K^\omega(\tilde{\phi}) \rangle_{G} = \langle \phi, \tilde{\phi} \rangle_{SU(2)}$. Since the norm of $\phi$ is zero, $\phi=0$. Therefore, $K^\omega$ is injective.
\end{proof}
This proposition implies that if the embedding map $K^\omega$ is isometric then $\tilde{P}$ exists\footnote{It is important to note that the injectivity is weaker condition than the isometry, thus the converse of the proposition is not true; injectivity does not guarantee isometry.}.

We are interested in whether the isometry of $K^\omega$ can be compatible with the restriction of the projection map being an inverse of $K^\omega$. The following proposition shows that they can be compatible under certain conditions, when $\tilde{P}$ is chosen to be the restriction of the projection map.

\prop{If $K^\omega$ is isometric and $S^\omega$ is an orthogonal projector then $\tilde{P}(G_e, g_e, x_v)=\overline{K(G_e, g_e, x_v)}$.}
\begin{proof}
Since $K^\omega$ is injective, $\tilde{P}$ exists. Consider the isometry condition for $K^\omega$
\begin{equation}
\langle K^\omega(\phi), K^\omega(\tilde{\phi}) \rangle_{G} = \langle \phi, \tilde{\phi} \rangle_{SU(2)}.
\end{equation}
The left-hand side of the condition reads
\begin{equation}
\int_{Q}[dx_v]^V\int_G [dG_e]^E \int_{SU(2)}[dg_edh_e]^E \overline{K^\omega(G_e, g_e, x_v)\phi(g_i)}K^\omega(G_e, h_e, x_v)\tilde{\phi}(h_i),
\end{equation}
and the right-hand side of the condition is
\begin{equation}
\int_{SU(2)} [dg_edh_e]^E\overline{\phi(g_e)}\tilde{\phi}(h_e)\prod_{e=1}^E\delta_{SU(2)}(g_eh_e^{-1})\bigg\rvert_{\mathcal{K}^{SU(2)}_{kin}} \quad.
\end{equation}
By setting two integral expressions equal, one can obtain
\begin{align}
0= \int_{SU(2)} [dg_edh_e]^E \overline{\phi(g_e)} \tilde{\phi}(h_e) \Big[\prod_{e=1}^E\delta_{SU(2)}(g_eh_e^{-1})\bigg\rvert_{\mathcal{K}^{SU(2)}_{inv}}-\int_Q [dx_v]^V \int_G [dG_e]^E \overline{K^\omega(G_e, g_e, x_v)}K^\omega(G_e, h_e, x_v)\Big] \quad .
\end{align}
This relation holds for any $\phi$, $\tilde{\phi}$ $\in$ $\mathcal{K}^{SU(2)}_{kin}$, thus
\begin{equation}\label{isogroup}
\int_Q [dx_v]^V \int_G  [dG_i]^E \overline{K^\omega(G_e, g_e, x_v)}K^\omega(G_e, h_e, x_v)=\prod_{e=1}^E\delta_{SU(2)}(g_eh_e^{-1})\bigg\rvert_{\mathcal{K}^{SU(2)}_{kin}} \quad .
\end{equation}
On the other hand, from the fact that $\tilde{P}$ is the inverse of $K^\omega$, it follows that
\begin{equation}
\phi(g_e)= \int_Q [dx_v]^V \int_{SU(2)} [dh_e]^E \int_G [dG_e]^E \tilde{P}(G_e, g_e, x_v)K^\omega(G_e, h_e, x_v)\phi(h_e) \quad .
\end{equation}
This implies
\begin{equation}\label{inverse1group}
\int_Q [dx_v]^V \int_G  [dG_e]^E \tilde{P}(G_e, h_e, x_v)K^\omega(G_e, g_e, x_v)=\prod_{e=1}^E\delta_{SU(2)}(g_eh_e^{-1})\bigg\rvert_{\mathcal{K}^{SU(2)}_{kin}}.
\end{equation}
Note that the right-hand side of equations (\ref{isogroup}) and (\ref{inverse1group}) are the same, from which it follows that 
\begin{equation}
\int_Q [dx_v]^V \int_G [dG_e]^E \Big(\overline{K^\omega(G_e,g_e, x_v)}-\tilde{P}(G_e, g_e, x_v)\Big)K^\omega(G_e,h_e, x_v)=0 \quad .
\end{equation}
We can conclude that $\tilde{P}(G_e, g_e, x_v)=K^\omega(G_e, g_e, x_v)$ up to a function $[K^\omega]^{\perp}(G_e, g_e)$ which is defined by the relation
\begin{equation}
\int_Q [dx_v]^V\int_G [dG_e]^E [K^\omega]^{\perp}(G_e, g_e, x_v)K^\omega(G_e,h_e, x_v)=0 \quad .
\end{equation}
In fact, $[K^\omega]^{\perp}(G_e, g_e, x_v)$ constitutes the part of the embedding map which does not satisfy the simplicity constraint. 

This contribution is then zero for any orthogonal projector type of $S^\omega$, therefore it does not play any role in $\tilde{P}$ because functions in the domain of $\tilde{P}$ always satisfy the simplicity constraint.
\end{proof}

\prop{If $K^\omega$ is isometric then $\overline{K^\omega(G_e, g_e, x_v)}$ is an integral kernel of $\tilde{P}$.}
\begin{proof}
By comparing (\ref{isogroup}) and (\ref{inverse1group}) in the proof of the previous proposition, one can conclude that $\overline{K^\omega(G_e, g_e, x_v)}$ plays the role of an integral kernel of $\tilde{P}$, the inverse map of $K^\omega$.
\end{proof}

\prop{If $\tilde{P}(G_e, g_e, x_v)=\overline{K^\omega(G_e, g_e, x_v)}$, then $K^\omega$ is isometric.}
\begin{proof}
\begin{align}
\langle K^\omega(\phi), K^\omega(\tilde{\phi}) \rangle_{G}&=\int_Q [dx_v]^V \int_G  [dG_e]^E \int _{SU(2)}[dg_edh_e]^E \overline{K^\omega(G_e, g_e, x_v)\phi(g_e)}K^\omega(G_e, h_e, x_v)\tilde{\phi}(h_e)\nonumber\\
&=\int_Q [dx_v]^V \int_G  [dG_e]^E \int_{SU(2)}[dg_edh_e]^E \tilde{P}(G_e, g_e, x_v)\overline{\phi(g_e)}K^\omega(G_e, h_e, x_v)\tilde{\phi}(h_e)\nonumber\\
&=\int_{SU(2)} [dg_e]^E \overline{\phi (g_e)}\tilde{\phi }(g_e) = \langle \phi, \tilde{\phi} \rangle_{SU(2)}.
\end{align}
\end{proof}

\prop{If $\tilde{P}(G_e, g_e, x_v)=\overline{K^\omega(G_e, g_e, x_v)}$, then $\tilde{P}$ is isometric.}
\begin{proof}
\begin{align}
\langle \tilde{P}(\psi^\omega), \tilde{P}(\tilde{\psi}^\omega) \rangle_{SU(2)}&= \int_Q [dx_vd\tilde{x}_v]^V\int_{SU(2)} [dg_e]^E \int_G [dG_e dH_e]^4 \overline{\tilde{P}(G_e, g_e, x_v)\psi^\omega(G_e, x_v)}\tilde{P}(H_e, g_e, \tilde{x}_v)\tilde{\psi}^\omega(H_i, \tilde{x}_v)\nonumber\\
&= \int_Q [dx_vd\tilde{x}_v]^V \int_{SU(2)} [dg_e]^E \int_G[dG_e dH_e]^E K^\omega(G_e, g_e, x_v) \overline{\psi^\omega(G_e, x_v)}\tilde{P}(H_e, g_e, \tilde{x}_v)\tilde{\psi}^\omega(H_e, \tilde{x}_v)\nonumber\\
&=\int_Q [dx_v]^V \int_G [dG_e]^E  \overline{\psi^\omega(G_e, x_v)}\tilde{\psi}^\omega(G_e, x_v) =\langle \psi^\omega, \tilde{\psi}^\omega \rangle_{G}
\end{align}
\end{proof}

\noindent To summarize, the injectivity of $K^\omega$ allows us to construct its inverse $\tilde{P}$. Moreover, if $K^\omega$ is isometric and $S^\omega$ is an orthogonal projector, then the integral kernels of $\tilde{P}$ and $K$ are complex conjugates to each other, which implies that $\tilde{P}$ is isometric. For a non-orthogonal projector $S^\omega$, $\tilde{P}$ is still isometric when the integral kernels of $\tilde{P}$ and $K$ are complex conjugates to each other.

\noindent For an orthogonal projector $S^\omega$, if $\tilde{P}$ is chosen to be the (restricted) projection map, then $K^\omega$ can be isometric if and only if $\tilde{P}(G_e,g_e, x_v) =\overline{K^\omega(G_e,g_e, x_v)}$. For a non-orthogonal projector $S^\omega$, $\tilde{P}(G_e,g_e, x_v) =\overline{K^\omega(G_e,g_e, x_v)}$ implies the isometry of $K^\omega$ and $\tilde{P}$. 

\noindent These results suggests a systematic way of finding two maps which are inverses to each other and isometric at the same time:\\
\indent \textit{1. For a given simplicity constraint, check the injectivity of the embedding map.}\\
\indent \textit{2. Once the embedding map is found to be injective, then rescale it in such a way that the map is isometric.}\\
\indent \textit{3. Define the restriction of the projection map such that the complex conjugate of the corresponding integral kernel coincides with the integral kernel of the embedding map.}\\

\section{The Dupuis-Livine type of maps}\label{section: DL map}
In section \ref{section:embedding and projection maps}, we have shown the symmetry requirement on the integral kernel implies the convolution between a $G$ character $\Theta$ and an $SU(2)$ character $\chi$. This same type of (gauge-fixed) embedding map, in the case of the EPRL imposition of simplicity constraints, has been defined and studied in \cite{PhysRevD.82.064044} by M. Dupuis and E. Livine. In that work, the projection map is defined as a restriction of $G$ to the $SU(2)$ subgroup. This definition of Dupuis-Livine (DL) embedding map and projection map has been then shown to lead to be the incompatibility of the requirement of isometry for the embedding map with the simultaneous requirement that the projection map is the (restriction of) inverse of the embedding map. 

In what follows, we consider the same type of embedding map, i.e. with the same symmetry requirements, but generalized to any type of the simplicity constraints. We investigate injectivity and isometry of the map so generalised. Furthermore, using the properties discussed in the section \ref{section:Relations}, we discuss which types of the simplicity constraints can provide a projection map that is an inverse of corresponding embedding map and simultaneously allows isometry of the embedding map.

The DL-type embedding maps are maps between kinematical, i.e. gauge invariant Hilbert spaces, thus one can gauge fix and then disregard the dependency on normal vectors for the $G$-dependent cylindrical functions. Moreover, we adopt a simple regularization for the inner product, especially in the Lorentzian theory, defining it up to the volume of the divergent integral over normal vectors.

\subsection{Maps in Riemannian theory}
In \cite{PhysRevD.82.064044} only Lorentzian theory were discussed. Here we extend their discussion defining the DL-type embedding map also for the Riemannian theory. Consider the embedding map defined as
\begin{equation}
\varphi^\omega(G_i)=\sum_{j_i^+, j_i^-, j_i}\prod_{i=1}^4 \Delta_{j_i}^{(j^+_i, j^-_i)}\int_{SU(2)}[dh_idg_i]^4\phi(g_i) \prod_{i=1}^4\Big(\chi^{j_i}(h_ig_i)\Theta^{(j_i^+, j_i^-)}(G_ih_i)\omega(j^+_i,j^-_i,j^-_i,\gamma_i)\Big) \quad .
\end{equation}
The map is defined on a single 4-valent vertex and its generalization to arbitrary graphs can be obtained by proper gluing procedure. The $\Delta$ factors in the maps are to be determined, depending on the properties we want the maps to have.

The integral kernel of the embedding map is
\begin{equation}
K^\omega(G_i, g_i, \mathbbm{1}) = \sum_{j_i^+, j_i^-, j_i}\prod_{i=1}^4 \Delta_{j_i}^{(j^+_i, j^-_i)}\int_{SU(2)} [dh_i]^4 \prod_{i=1}^4\Big(\chi^{j_i}(h_ig_i)\Theta^{(j_i^+, j_i^-)}(G_ih_i)\omega(j^+_i,j^-_i,j^-_i,\gamma)\Big) \quad .
\end{equation}
The spin representation of the integral kernel is
\begin{equation}
[K^\omega(\mathbbm{1})]^{j^+_ij^-_ij_i}_{m_i^+n_i^+m_i^-n_i^-m_in_i}=\prod_{i=1}^4\Big(\frac{\Delta^{(j^+_i,j^-_i)}_{j_i}}{d_{j_i}}\overline{C^{j^+_ij^-_ij_i}_{m^+_im^-_im_i}}C^{j^+_ij^-_ij_i}_{n^+_in^-_in_i}\omega(j^+_i, j^-_i, j_i, \gamma)\Big) \quad .
\end{equation}
As remarked, the properties of the map depend on the $\Delta$ factor and the simplicity constraint imposition encoded in the coefficient $\omega$. 

Consider an embedding map without the simplicity constraints, i.e. with $\omega=1$. If one chooses the $\Delta$ factor as
\begin{equation}\label{delta}
\Delta^{(j^+_i, j^-_i)}_{j_i} = \sqrt{d_{j^+_i}d_{j^-_i}d_{j_i}^3} \quad ,
\end{equation}
then the embedding map is injective and isometric.

Another interesting choice of $\Delta$ factor is
\begin{equation}
\Delta^{(j^+_i, j^-_i)}_{j_i} =d_{j_i}^2.
\end{equation}
This choice accounts for the case in which, when the domain of the $Spin(4)$ function are restricted to the $SU(2)$ subgroup, the function coincides with the $SU(2)$ function before embedding:
\begin{equation}\label{delta2}
\varphi^\omega(g_i)=K^\omega(\phi)(g_i)=\phi(g_i)
\end{equation}
for any $g\in SU(2)$ and with the EPRL simplicity constraint being encoded in the coefficient $\omega$. If the projection map is defined as a restriction of the domain of $Spin(4)$ function
\begin{equation}
P(\varphi^\omega)(g_i)=\phi(g_i) \quad,
\end{equation}
the (restriction of) projection map is the inverse of the embedding map, $P\circ K^\omega =id_{SU(2)}$. However, the choice (\ref{delta2}) does not yield an isometric embedding map. This incompatibility between the two properties stems from the fact that $\tilde{P}(G_e, g_e, x_v)\neq \overline{K^\omega(G_e, g_e, x_v)}$.
The systematic method suggested in the section \ref{section:Relations} shows that if one gives up the requirement concerning the restriction of $G$ to $SU(2)$, and defines the projection map in a non-trivial way, one can find the maps satisfying both conditions under the assumption that the embedding map is injective. In the following section, we will pursue this route, and investigate a generalised DL-type map with several different chocies of simplicity constraints imposition, focusing on the compatibility between isometry of the embedding map and the requirement that the projection map being an inverse of the embedding map.

Now consider an embedding map $K^\omega$ with non-trivial simplicity constraints. The embedding map $K^\omega$ is injective if 
\begin{equation}
\sum_{j_i}\prod_{i=1}^4\Big(\Delta^{(j^+_i, j^-_i)}_{j_i}\omega(j^+_i, j^-_i, j_i, \gamma)\Big)\phi^{j_i}_{m_in_i}=0  \text{ $\forall$ } j^+_i, j^-_i , m_i, n_i \text{ implies } \phi^{l_i}_{p_iq_i}= 0 \text{ $\forall$ }  l_i, p_i, q_i \quad .
\end{equation}
If the CG coefficients are invertible then $K^\omega$ is injective. Since the invertibility of CG coefficients depends on the simplicity constraint coefficient, injectivity of the embedding map also depends on the constraint imposition.
For an injective $K^\omega$, the isometry condition is then
\begin{equation}\label{eq:isometryF}
\sum_{j^{\pm}_i}\prod_{i=1}^4\Big(\frac{\big(\Delta^{(j^+_i, j^-_i)}_{j_i}\big)^2}{d_{j_i}^3d_{j_i^+}d_{j_i^-}}\omega^2(j^+_i, j^-_i, j_i, \gamma)\Big)=1 \quad .
\end{equation}
The isometry condition also depends on both $\Delta$ factor and the simplicity constraint coefficient. 
The injectivity and the isometry of the map with non-trivial simplicity constraints are treated in the next section.

\subsection{Maps in Lorentzian theory}
Consider the embedding map defined as
\begin{equation}
\varphi^\omega(G_i)=\sum_{a_i, j_i}\int \prod_{i=1}^4 \Big(\mu(\rho_i, a_i) d\rho_i \Big)\prod_{i=1}^4 \Delta_{j_i}^{(\rho_i, a_i)} \int_{SU(2)}[dh_idg_i]^4 \phi(g_i)\prod_{i=1}^4\Big(\chi^{j_i}(h_ig_i)\Theta^{(\rho_i, a_i)}(G_ih_i)\omega((\rho_i, a_i), j_i,\gamma_i)\Big) \quad .
\end{equation}
Like in the Riemannian theory, the map is defined on a single 4-valent vertex. 
The integral kernel of $K^\omega$ and its spin representation are
\begin{align}
K^\omega(G_i, g_i, \mathbbm{1}) &= \sum_{a_i, j_i}\int \prod_{i=1}^4 \Big(\mu(\rho_i, a_i) d\rho_i \Big)\prod_{i=1}^4 \Delta_{j_i}^{(\rho_i, a_i)} \int_{SU(2)} [dh_i]^4 \prod_{i=1}^4\Big(\chi^{j_i}(h_ig_i)\Theta^{(\rho_i, a_i)}(G_ih_i)\omega((\rho_i, a_i), j_i,\gamma_i)\Big), \\
[K^\omega(\mathbbm{1})]^{(\rho_i, a_i)k_i}_{j_im_il_in_ip_iq_i}&=
\begin{cases}
\prod\limits_{i = 1}^{4}\Big(\frac{\Delta_{k_i}^{(\rho_i, a_i)}}{d_{k_i}}\delta_{p_im_i}\delta_{q_in_i}\delta^{l_ij_i}\delta^{k_il_i}\omega((\rho_i, a_i), k_i, \gamma)\Big) &\text{if $l_i \in a+\mathbb{N}$ } \\
0 &\text{if $l_i \notin a+\mathbb{N}$ \qquad .}
\end{cases}
\end{align}

Consider an embedding map without the simplicity constraint, i.e. with $\omega=1$. The map is isometric if
\begin{equation}
\sum_{a_i}\int \prod_{i=1}^4\Big(\mu(\rho_i, a_i)d\rho_i\Big) \prod_{i=1}^4\Big(\Delta^{(\rho_i, a_i)}_{k_i}\Big)^2 = \prod_{i=1}^4 d_{k_i}.
\end{equation}
Such $\Delta$ factor is necessarily distributional in the Lorentzian theory.

Now consider an embedding map $K^\omega$ with a non-trivial simplicity constraint coefficient. The map is injective if 
\begin{equation}
\sum_{j_i}\Delta^{(\rho_i, a_i)}_{j_i}\tilde{\omega}((\rho_i, a_i), j_i, \gamma)\phi^{j_i}_{m_in_i}=0 \text{ $\forall$}\rho_i, a_i, m_i, n_i\text{ implies } \phi^{l_i}_{p_iq_i}=0 \text{ $\forall$}l_i, p_i, q_i \quad ,
\end{equation}
and isometric if 
\begin{equation}
\sum_{a_i}\int \prod_{i=1}^4 \Big( \mu(\rho_i, a_i)\big(\omega((\rho_i, a_i), k_i, \gamma)\big)^2 d\rho_i \Big) \prod_{i=1}^4(\Delta_{k_i}^{(\rho_i,a_i)})^2=\prod_{i=1}^4d_{k_i} \quad .
\end{equation}
The injectivity and isometry of the maps depend again on the explicit form of the simplicity constraint coefficient $\omega$.

Another interesting choice of the $\Delta$ factor is when
\begin{equation}\label{delta3}
\Delta_{k_i}^{(\rho_i, a_i)} = \frac{d_{k_i}}{\mu(\rho_i, a_i)} \quad .
\end{equation}
This choice corresponds to the case in which, when the domain of the $SL(2, \mathbb{C})$ cylindrical function is restricted to the $SU(2)$ subgroup, the function coincides with the original $SU(2)$ cylindrical function:
\begin{equation}
\varphi^\omega(g_i) = K^\omega(\phi)(g_i)= \phi(g_i)
\end{equation}
for any $g \in SU(2)$, and with the EPRL simplicity constraint having been imposed (and under certain restriction on the domain of $\mathcal{K}^{SU(2)}_{kin}$ such that the embedding map is injective). If the projection map $P$ is defined as a restriction of the domain (as in \cite{PhysRevD.82.064044})
\begin{equation}
P(\varphi)(g_i)=\phi(g_i),
\end{equation}
then the (restriction of) projection map is an inverse of the embedding map $P\circ K^\omega = id_{SU(2)}$. However, the choice (\ref{delta3}) does not give an isometric embedding map. This incompatibility between these two properties stems from the fact that $\tilde{P}(G_e, g_e, x_v)\neq \overline{K^\omega(G_e, g_e, x_v)}$. Also, in the Lorentzian case, the systematic method suggested in section \ref{section:Relations} can resolve the incompatibility, which is what we discuss in the following section.

\section{Case studies}\label{section:Case studies}
In section \ref{section:embedding and projection maps} we have shown that the embedding map between kinematical, gauge-invariant Hilbert spaces is a convolution of characters. This form of embedding map has been considered then in section \ref{section: DL map}, analyzing its generic properties. In this section, we investigate the properties of the embedding map with different choices of simplicity constraints imposition.

\subsection{Riemannian BC model}
The Riemannian Barrett-Crane (BC) model \cite{doi:10.1063/1.532254}imposes the simplicity constraint strongly in the generalised spin network basis: the simplicity constraint written in terms of two Casmir operators of $Spin(4)$ is set to zero, in fact corresponding to the vanishing of one of them. This model can be seen as a limiting case $\gamma \rightarrow \infty$ of the EPRL model \cite{ENGLE2008136} and of the BO and BO-Duflo models \cite{BO2, Finocchiaro2019}. The fusion coefficient of the model is
\begin{equation}
\omega_{BC}(j^+, j^-, j, \gamma=\infty)= \delta^{j^+j^-}\delta^{j0} \quad .
\end{equation}
It is immediate to verify that this imposition of the simplicity constraints gives a clearly non-injective (and not thus isometric) embedding map. 

This is not surprising since the simplicity constraints totally trivialize the dependence on the data from the rotation $SU(2)$ subgroup of $Spin(4)$. Let us point out, however, that this does not mean necessarily that no map between covariant states and $SU(2)$-based states can be defined. In fact, we expect such map to be possible, since the equivalence of possible descriptions of simplicial geometry in terms of both covariant and $SU(2)$ data remains true also in this case (i.e. in absence of the Immirzi paramemeter). It implies, however, that the relevant $SU(2)$ cannot be identified with the rotation subgroup of $Spin(4)$, as we have assumed in this work.

\subsection{Riemannian EPRL model}
The Engle-Pereira-Rovelli-Livine (EPRL) model \cite{ENGLE2008136} encodes the simplicity constraints with finite Immirzi parameter and results from a weak imposition of them, as necessary in this case, and using a Master constraint technique: the simplicity constraints, expressed in terms of Casimir operators, are squared and then minimized when imposed on quantum states (for details on the imposition, see \cite{Perez2013}). The corresponding EPRL fusion coefficient is
\begin{equation}
\omega_{EPRL}(j^+, j^-, j, \gamma)= \delta^{j^+,|\frac{1+\gamma}{1-\gamma}|j^-}\delta^{j, |\frac{2}{1+\gamma}|j^+} \quad.
\end{equation}
Since $j^{\pm}$ label $SU(2)$ representations, $j^{\pm}=|(1\pm \gamma)|j/2$ should be a non-negative half-integer $\mathbb{N}_0/2$. This requires $\gamma$ to be a rational number. For given values of $\gamma$ and $j$, however $|(1\pm \gamma)|j/2$ is not always in $\mathbb{N}_0/2$. For such $j$, the injectivity condition is not satisfied. Thus, the embedding map with the EPRL imposition of simplicity constraints is not injective in general.

However, if $\gamma$ is any odd integer, $j^+$ and $j^-$ become non-negative half-integesr. In this case, the map is injective. As the simplicity constraint coefficient $\omega$ has unit norm, the EPRL embedding map is isometric when one chooses the $\Delta$ factor to be the one given in (\ref{delta}). Moreover, if one defines the projection map such that its integral kernel is the complex conjugate of the integral kernel of the embedding map, the restriction of the projection map is the inverse of the embedding map, which is isometric.

\subsection{Riemannian FK model}
The Freidel-Krasnov (FK) spin foam model employs a  decomposition of cylindrical functions in terms of group coherent states (for each irreducible representation space, thus eigenspace of the Casimir operators), to impose the simplicity constraints, in their linear version \cite{Gielen2010}, on coherent state parameters. These are in fact interpreted as quantum counterpart of the bivector variables which are subject to simplicity constraints at the classical level \cite{Freidel2008}.
The FK model coincides with the EPRL model when $\gamma < 1$, up to some ambiguities in the fusion coefficient, but with the same restriction on the group representations appearing in the decomposition \cite{Finocchiaro2019}. The group representations are related as
\begin{align}
\gamma < 1,\hspace{0.2cm} j^{\pm}&=(1\pm\gamma) j/2, \\
\gamma > 1,\hspace{0.2cm} j^{\pm}&=(\gamma\pm1)j/2 \quad 
\end{align}
and the coefficients $\omega$ in each case are
\begin{align}
\gamma < 1,\hspace{0.2cm} \omega_{FK}^{\gamma < 1}&= \delta^{j^+, (1+\gamma)j/2}\delta^{j^-, (1-\gamma)j/2}|C^{j^+j^-j}_{j^+j^-j}|,\\
\gamma > 1,\hspace{0.2cm} \omega_{FK}^{\gamma > 1}&= \delta^{j^+, (\gamma+1)j/2} \delta^{j^-, (\gamma-1)j/2}|C^{j^+j^-j}_{j^+-j^-j}| \quad .
\end{align}
For the same reason as in the EPRL case, the FK model is not injective in general.

If $\gamma$ is any odd integer, $j^+$ and $j^-$ become non-negative half-integers. In this case, the map is injective. As the simplicity constraint $\omega$ is not normalized to one, a different expression for $\Delta$ has to be chosen for achieving the isometry of the embedding map. Under the rescaling of the $\Delta$ factor given in (\ref{delta})
\begin{equation}
\Delta^{(j^+_i,j^-_i)}_{j_i} \rightarrow \frac{\Delta^{(j^+_i,j^-_i)}_{j_i}}{|C^{j^+j^-j}_{j^+\pm j^-j}|} \quad ,
\end{equation}
the FK embedding map is isometric, which can be justified by (\ref{eq:isometryF}). As a result, one can also construct a projection map whose restriction is the inverse of the isometric embedding map.

\subsection{Riemannian BO-Duflo model}
The Baratin-Oriti (BO) model \cite{BO1, BO2} is constructed from non-commutative metric/flux formulation of spin foam models and also imposes the linear version of simplicity constraints \cite{Gielen2010} directly on such flux variables at the quantum level, since these are the closest quantum translation of the bivector variables of the discrete classical theory. 
%Unlike group field theory where group manifolds form a configuration space, the model chooses (several copies of) Lie algebras as a configuration space that is a non-commutative geometry. 
The construction of quantum theory starting from the discrete classical one requires of course a choice of quantization map for the classical variables. The quantization map for the original BO model is the FLM map (used in \cite{BO1,Oriti2014} and first introduced in \cite{Freidel2008}), but a more recent variation of the construction adopts the Duflo map. This has nicer mathematical properties \cite{ASENS_1977_4_10_2_265_0} (see also \cite{Guedes_2013}) and wider applicability to any semi-simple and locally compact group, but it also simplifies computations, providing a new spin foam model for whose fusion coefficients one can obtain an explicit, complicated but manageable expression \cite{Finocchiaro2019}.

The simplicity constraint coefficient of the BO-Duflo model \cite{Finocchiaro2019} is
\begin{equation}\label{BO_omega}
\omega_{BO}(j^+, j^-, j, \gamma) = \frac{(-1)^{j^++j^-+j}}{\pi\sqrt{(2j^++1)(2j^-+1)}} \sum_{a=0}^{\lambda}(\text{sign}(\beta))^a \Big\{
\begin{matrix}
a & j^- & j^-\\
j & j^+ & j^+
\end{matrix}\Big\}\mathcal{T}_a^{j^+j^-}(|\beta|) \quad ,
\end{equation}
where $\lambda=2\text{min}(j^+,j^-)$ and $\beta=\frac{\gamma-1}{\gamma+1}$. The function $\mathcal{T}$ is 
\begin{equation}
\mathcal{T}_a^{j^+j^-}(|\beta|)=(-1)^a(-1+1)\int^{2\pi}_0 d\theta \Omega(\beta, \theta) \sin^2 \frac{\theta}{2} \chi^{j^+}_a(\theta_\beta) \chi^{j^-}_a(\theta)
\end{equation}
where $\theta_\beta= |\beta|\theta$ and the generalized character $\chi^j_a(\theta)$ of $SU(2)$ representations and $\Omega$ are defined as
\begin{equation}
\chi^j_a(\theta)=i^a \sum_{p=-j}^j e^{-ip\theta}C^{ajj}_{0pp}, \indent \Omega(\beta, \theta) = \frac{\sin\frac{|\beta|\theta}{2}}{|\beta|\sin \frac{\theta}{2}} \quad .
\end{equation}
It is not easy to check the injectivity and isometry of the map directly from (\ref{BO_omega}), i.e. at the fully analytic level, due to the complexity of the expression of the coefficient $\omega$. However, we can study the behaviour of the coefficient numerically and also check some limiting cases of the model (also discussed in \cite{Finocchiaro2019}).

The structure of the fusion coefficient $\omega$ as well as numerical plots show a simple power-low behavior, for $\beta>0$, for the function $\Pi$ on the right hand side of the following formula:
\begin{equation}\label{numericalsum}
\sum_{j^-j^+}d^a_{j^-}d^a_{j^+}\omega^2(j^+, j^-, j, \gamma(\beta)) = \Pi(a, \beta, j)
\end{equation}
where $\Pi(a,\beta, j)=c(a,\beta)j^{\eta(a,\beta)}$ for some coefficients $c(a,\beta)$ and $\eta(a,\beta)$. The isometry of the BO-Duflo model can be investigated from this numerically derived formula.  For a positive $\beta$, the isometry condition can always be fulfilled when the domain of map is restricted to $j\neq0$ because then
\begin{equation}
\sum_{j^-j^+}\frac{d^a_{j^-}d^a_{j^+}}{c(a, \beta)j^{\eta(a,\beta)}}\omega^2(j^+, j^-, j, \beta) = 1
\end{equation}
which implies
\begin{equation}
\Delta^{(j^+,j^-)}_j=\pm \sqrt{\frac{d_j^3d^{a+1}_{j^-}d^{a+1}_{j^+}}{c(a,\beta) j^{\eta(a,\beta)}}}
\end{equation}
satisfies the isometry condition. The indicated restriction of the range of allowed representation labels is reasonable (even if it is not automatically implemented by the amplitudes of the model), since it coincides with one of the requirements for cylindrical consistency in canonical LQG. 

Using the linear-regression method, we can also numerically determine $\Pi$ for any $\beta>0$ (thus, $\gamma > 1$):
\begin{eqnarray}
&&\Pi(a, \beta, j)= c j^\eta, \eta=\eta(a,\beta)= -0.98519 + 1.96698 a - 0.000917856 \beta,\nonumber\\
&&k(a,\beta) = log\text{ }c(a,\beta) = 1.98163 + 1.33946 a + 0.0749463 a^2 - 6.44261 \beta + 9.22207 \beta^2,
\end{eqnarray}
where the standard error, t-statistic, and P-value of each estimated coefficient for $k$ and $\eta$ are given in Table I and Table II.
\begin{table}[h!]
\begin{center}
\begin{tabular}{c|c|c|c|c}
 & \text{Estimate} & \text{Standard Error} & \text{t-Statistic} & \text{P-Value} \\
\hline
\text{constant term} & $1.98163$ & $1.19506$ & $1.65819$ & $0.10225$ \\
\text{coefficient of $a$} & $1.33946$ & $0.190011$ & $7.0494$ & $1.639056269781041 \times 10^{-9}$ \\
\text{coefficient of $\beta$} & $-6.44261$ & $6.79493$ & $-0.948149$ & $0.346677$ \\
\text{coefficient of $a^2$} & $0.0749463$ & $0.0832435$ & $0.900326$ & $0.371376$ \\
\text{coefficient of $\beta^2$} & $9.22207$ & $7.77542$ & $1.18605$ & $0.240055$ \\
\end{tabular}
\caption{Numerical estimation for $k(a,\beta)$}
\label{tab: Table 1}
\end{center}
\end{table}
\begin{table}[h!]
\begin{center}
\begin{tabular}{c|c|c|c|c}
& Estimate & Standard Error & t-Statistic & P-Value\\
\hline
\text{constant term} & $-0.98519$ & $0.00726067$ & $-135.689$ & $8.75140531349814\times 10^{-65}$\\
\text{coefficient of $a$} & $1.96698$ & $0.00257899$ & $762.697$ & $1.6915975668819295 \times 10^{-101}$\\
\text{coefficient of $\beta$} & $-0.000917856$ & $0.0151906$ & $-0.0604226$ & $0.952065$ \\
\end{tabular}
\caption{Numerical estimation for $\eta(a,\beta)$}
\label{tab: Table 2}
\end{center}
\end{table}

These numerical analysis shows $\eta(a,\beta)$ (for $\beta>0$ regime) is very weakly dependent on $\beta$ (almost constant) and has a simple expression in terms of $\eta(a, \beta)=2a-1$. Having accepted this expression for $\eta$, $\Pi$ becomes independent of $j$ when $a=1/2$:
\begin{equation}
\sum_{j^-j^+}\sqrt{d_{j^-}d_{j^+}}\omega^2(j^+, j^-, j, \beta) = c(\frac{1}{2},\beta).
\end{equation}
This relation shows $\Delta$ of the form 
\begin{equation}
\Delta^{(j^+,j^-)}_{j}= \pm \frac{d^{3/2}_jd^{3/4}_{j^+}d^{3/4}_{j^-}}{\sqrt{c(\frac{1}{2},\frac{3}{4})}}
\end{equation}
satisfies the isometry condition (\ref{eq:isometryF}) even without restricting the domain to $j\neq0$.

For a negative $\beta$, the numerical analysis suggests $\Pi(a, \beta, j)$ on the right hand side of (\ref{numericalsum}) follows a modulated power-law behaviour instead of a simple power law
\begin{equation}
\Pi(a, \beta, j) = \frac{k_1(a,\beta) \text{sin} j-k_2(a,\beta)}{j^{\eta(a,\beta)}}
\end{equation}
with some parameter coefficients $k_1(a,\beta)$, $k_2(a,\beta)$, and $\eta(a,\beta)$. The isometry condition can be satisfied for different values of negative $\beta$, but it can be checked (numerically) only on a case by case basis because fits are less accurate than positive $\beta$ case.

Now let us consider some limiting cases of the BO-Duflo model.
\subsubsection*{$\beta \rightarrow1$: the BC model}
In the limit $\beta\rightarrow1$ (the Immirzi parameter $\gamma$ goes to infinity) the simplicity constraints coefficient reduces to the one of the BC model
\begin{equation}
\omega_{BO}(j^+, j^-, j, \infty)=\delta^{j^+j^-}\delta^{j0}
\end{equation}
As discussed before, the resulting embedding map cannot be injective.
\subsubsection*{$\beta \rightarrow-1$: the topological model}
In the limit $\beta\rightarrow -1$, and $\gamma$ goes to zero, meaning that one expects the Holst term to be dominant over the geometrical term in the Palatini-Holst classical action. This could be then a regime in which the theory becomes topological (but see \cite{Benedetti:2011nd} for a careful analysis at the perturbative quantum level of the continuum theory, indicating a more subtle outcome). The limiting simplicity constraint coefficient becomes
\begin{equation}
\omega_{BO}(j^+, j^-, j, 0)=\frac{(-1)^{j^++2j^-}}{(2j^-+1)}\delta^{j^+ j^-}\{j^+, j^-, j\}
\end{equation}
where $\{j^+, j^-, j\} = 1$ if three $j$'s satisfy the triangle inequality and it vanishes instead if it does not. The constraint gives an injective embedding map. For the isometry of the embedding map, the $\Delta$ factor given in (\ref{delta}) has to be rescaled as follows
\begin{equation}
\Delta^{(j^+_i,j^-_i)}_{j_i} \rightarrow \Delta^{(j^+_i,j^-_i)}_{j_i}(-1)^{j^-_i}(2j^-_i+1).
\end{equation}
Under this choice, an isometric restricted projection map can be constructed, which is an inverse of the embedding map.

\subsubsection*{$\beta \rightarrow 0$: the Ooguri model}
In the limit $\beta\rightarrow 0$, the $\gamma$ converges to one. This model seems to correspond to the $SU(2)$ Ooguri model for topological BF theory \cite{Ooguri1992}. The fusion coefficient is
\begin{equation}
\omega_{BO}(j^+, j^-, j, 1)=\frac{2(-1)^{2j^-}}{(2j^-+1)^2}
\end{equation}
The constraint gives an injective embedding map. Under the rescaling of $\Delta$,
\begin{equation}
\Delta^{(j^+_i,j^-_i)}_{j_i} \rightarrow \frac{(-1)^{2 j^-_i}}{2}(2j_i+1)^2\Delta^{(j^+_i,j^-_i)}_{j_i},
\end{equation}
the embedding map is isometric and one can construct the restricted projection map which is isometric as well as an inverse of the embedding map.

\subsection{Lorentzian BC model}
The fusion coefficient encoding the simplicity constraint in the Lorentzian BC model \cite{Barrett2000} is 
\begin{equation}
\omega_{BC}((\rho, a), j)= \delta^{a0}\delta^{j0}
\end{equation}
where the Immirzi parameter $\gamma$ is absent. Like the Riemannian case, the injectivity condition is not satisfied, for the same structural reason. The same comment about the possibility of an alternative definition of the quantum geometric maps applies too.

\subsection{Lorentzian EPRL model}
The fusion coefficient encoding the simplicity constraint in the Lorentzian EPRL model is given by
\begin{equation}
\omega_{EPRL}((\rho, a), j, \gamma)= \delta^{j, a} \delta(\rho-\gamma a).
\end{equation}
This coefficient corresponds to an embedding map that does not satisfy the injectivity condition because it does not include $j_i=0$. Two $SU(2)$ cylindrical functions whose spin representations are different only at $j_i=0$ (at least for one $i$ among the possible four) are mapped to the same $SL(2,\mathbb{C})$ cylindrical function. However, if one restricts the domain of the embedding map to the $SU(2)$ functions which do not contain any $j_i=0$ contributions then the embedding map is injective. In fact, as we have remarked for the BO-Duflo model, this restriction of the domain can be understood as part of the requirements for cylindrical consistency. 

The inner product between two functions satisfying the simplicity constraint diverges, if naively defined, because the simplicity constraint operator is applied twice, with each application projecting out a non-compact part of the domain. In fact, any simplicity constraint whose implementation has this type of effect (as we may expect for other Lorentzian models) would produce a similar divergence. This divergence can be regularized by simply dropping the redundant delta distribution coming from the second imposition of the simplicity constraint.

Under the restriction of the domain which enforces the injectivity of the embedding map, the isometry condition for this model can be achieved with the $\Delta$ factor
\begin{equation}\label{Delta2}
\Delta_{k_i}^{(\rho_i, a_i)}= \sqrt{\frac{d_{k_i}}{\mu(\rho_i, a_i)}}.
\end{equation}
The method suggested in the section \ref{section:Relations} requires additional regularization to be implemented, due to the already pointed out singular feature of the Lorentzian EPRL imposition: once the projection map is defined such that $K^\omega(G_i, g_i, x_v)=\overline{P(G_i, g_i, x_v)}$, its image always carries a redundant $\delta(0)$. After dropping one such $\delta(0)$ as a regularization, one can achieve $P\circ K^\omega=id_{SU(2)}$.

\section{Conclusion}
We have defined quantum geometric maps between $SU(2)$ quantum states of geometry, as used in the canonical loop quantum gravity context, and covariant $SL(2, \mathbb{C})$-based quantum states of geometry, as naturally arising from the quantization of simplicial geometry in the context of spin foam models, following the formulation of gravity as a constrained topological BF theory. In doing so, we generalised existing work in the spin foam literature. In particular, we provided a definition which does not depend on any specific choice for the imposition of the simplicity constraints (the ones leading from topological BF theory to gravity, in such a way that our results applies to all current spin foam (and group field theory) models of quantum geometry.  In this general setting, we have analysed the properties of such quantum geometric maps and the mutual relations and compatibilities between them, as well as the role of the specific strategies for the imposition of the simplicity constraints.

We have shown that requiring the usual gauge symmetries on the domain and target space of the maps produces for the embedding map the same convolution structure that had been assumed for the DL embedding map. These DL-type maps, by construction, show a generic incompatibility between the requirement of isometry and the desired property that embedding and projection map are inverse to each other (under the restriction of the projection map), if the projection map is simply defined as a restriction of the domain of $SL(2,\mathbb{C})$ cylindrical functions to $SU(2)$. However, our general analysis shows that, if one drops this last assumption, one can always find a pair of embedding and projection maps, of the same DL-type, satisfying the two previously incompatible properties. We provide a simple algorithmic procedure, as well as the required conditions on the simplicity constraints imposition, for such reconciliation. 

We have also analysed what our conditions imply for existing spin foam models, based on different impositions of the simplicity constraints. In particular, we have shown that embedding maps of the DL type can be an isometry and an inverse of (the restriction of) the projection map for the Riemannian EPRL-FK models with an odd integer $\gamma$, the Riemannian BO-Duflo model, on the basis of a numerical analysis, and for the Lorentzian EPRL model (with proper regularization). 

These results improve our understanding of the quantum geometry underlying spin foam models and group field theories for 4-dimensional quantum gravity, and of the imposition of simplicity constraints that underlies them. It also contributes to clarifying the desired connection between the same models of quantum geoemtry and the description of the same arising from canonical Loop Quantum Gravity.

\bibliographystyle{unsrt}
\bibliography{embedding.bib}

\end{document}